\documentclass{article}%
\usepackage{amsmath}
\usepackage{amsfonts}
\usepackage{amssymb}
\usepackage{graphicx}%
\providecommand{\U}[1]{\protect\rule{.1in}{.1in}}
\newtheorem{theorem}{Theorem}

\begin{document}

\title{Placing hidden properties of quantum field theory into the forefront:\\wedge localization and a critical look at past S-matrix attempts \\{\small To the memory of Hans-J\"{u}rgen Borchers (1926-2011)}}
\author{Bert Schroer\\present address: CBPF, Rua Dr. Xavier Sigaud 150, \\22290-180 Rio de Janeiro, Brazil\\email schroer@cbpf.br\\permanent address: Institut f\"{u}r Theoretische Physik\\FU-Berlin, Arnimallee 14, 14195 Berlin, Germany}
\date{November 2012\ }
\maketitle

\begin{abstract}
Recent progress about "modular localization" reveals that, as a result of the
S-Matrix in its role of a "relative modular invariant of wedge-localization,
one obtains a new non-perturbative constructive setting of local quantum
physicis which only uses intrinsic (independent of quantization) properties.
The main point is a derivation of the particle crossing property from the KMS
identity of wedge-localized subalgebras in which the connection of
incoming/outgoing particles with interacting fields is achieved by "emulation"
of free wedge-localized fields within the wedge-localized interacting algebra.
The suspicion that the duality of the meromorphic functions, which appear in
the dual model, are not related with particle physics, but are rather the
result of Mellin-transforms of global operator-product expansions in conformal
QFT is thus confirmed. The connection of the wedge-localization setting with
the Zamolodchikov-Faddeev algebraic structure is pointed out and an Ansatz for
an extension to non-integrable models is presented.

Modular localization leads also to a widening of the renormalized perturbation
setting by allowing couplings of string-localized higher spin fields which
stay within the power-counting limit. This holds the promise of a Hilbert
space formulation which avoids the use of BRST Krein-spaces. 

.

\end{abstract}

\section{Introduction}

The course of quantum field theory (QFT) was to a large extend determined by
three important conceptual conquests: its 1926 discovery by Pascual Jordan in
the aftermath of what in recent times is often referred to as the
\textit{Einstein-Jordan conundrum} \cite{Du-Ja} \cite{E-J}, the discovery of
renormalized perturbation in the context of quantum electrodynamics (QED)
after world war II, and the nonperturbative insights into the particle-field
relation initiated in the Lehmann-Symanzik-Zimmermann (LSZ) work on scattering
theory, its derivation from first principles \cite{Haag}, as well as its
application to the rigorous derivation of the particle analog of the
Kramers-Kronig dispersion relations and the subsequent successful experimental
test of QFT's foundational causality principle. Their experimental
verification strengthened the trust in the causal localization principle of
QFT.\ The later gauge theory of the standard model resulted from an extension
of the ideas which already had led to QED. Besides some successes it led to
most of the still open problems of actual research.

Jordan's observations in his dispute with Einstein \cite{E-J} led to an
extension of quantization to matter waves, but its main point, the thermal
character of fluctuation in a vacuum state restricted to the local observables
of a subvolume which is needed did not receive the conceptual attention which,
being a characteristic property which distinguishes QFT from quantum mechanics
(QM), it would have deserved. It could have revealed itself as a
Gedankenexperiment of the kind which many decades later was proposed by Unruh
\cite{Unruh}\cite{Sewell}.

Both Gedankenexperiments demonstrate a thermal consequence of causal
localization whose early comprehension could have changed the path of QFT
history. When Jordan's incomplete calculation was published as a separate
section in the famous 1926 Dreim\"{a}nnerarbeit with Born and Heisenberg, his
coauthors had some reservations since it contained problematic aspects which
had no place in the previously discovered QM, but which they were not able to
clearly formulate and communicate.

Only several years later Heisenberg challenged Jordan in a letter about a
missing logarithmic term $\sim-ln\varepsilon$ in his calculation of the
fluctuation spectrum of the 2-dimensional chiral QFT which was Jordan's
favorite model\footnote{This chiral current model ("2-dim. photon") was the
only QFT which Jordan ever used in computations, in particular for his
"neutrino theory of light" which was nothing more than what nowdays is called
bosonization/fermionization \cite{Jo}. \ The prevalent incorrect idea was that
QFT like QM does not depend on spacetime dimensions in an essential way which
is only correct in QM but breaks down in QFT as soon as the spin $s\geq1~$(see
later).}, where $\varepsilon$ is a length which characterizes the "fuzzyness"
(deviation from sharp localization) at the endpoints of Jordan's localization
interval. This was the birth of Heisenberg's discovery of vacuum fluctuation
near the localization boundaries with $\varepsilon$ the "attenuation length"
("fuzzyness" of localization-boundary) conceded to the vacuum polarization cloud.

In his famous paper which he wrote after challenging Jordan, Heisenberg showed
that the localization of dimensionless quantum charges ("partial charge") in
QFT behaves quite different from partial charges in QM, for which localization
in the sense of Born is well-defined and complies with naive intuition. He
discovered that partial charges lead to a proportionality in terms of a
dimensionless area $area/\varepsilon^{2}$ which amounts to a quadratic
divergence in the sharp boundary limit $\varepsilon\rightarrow0$ (and to the
logarithmic divergence in Jordan's two-dimensional model). With the phenomenon
of vacuum polarization Heisenberg exposed one unexpected characteristic
consequence of causal localization which separates QFT from QM.

The other inexorable epiphenomenon of localization is "thermalization", i.e.
the fact that the restriction of the global pure vacuum state to observables
which are localized in a causally completed spacetime region turns it into an
impure state on all local measurements in that region; in fact this impure
state is thermal in the sense that it has the KMS property\footnote{In QM this
property only arises in the thermodynamic limit of Gibbs states. The issue of
heat bath thermality versus localization caused thermality is at the heart of
the Einstein-Jordan conundrum.} with respect to an intrinsically defined
"modular" Hamiltonian (section 3) which is not necessarily the same as that
measured by a thermometer or by the time which describes a Hamiltonian
evolution in some non-inertial system.

This knowledge was not accessible at the time of the Einstein-Jordan dispute,
but in retrospect it is clear that it would have been the most important
conceptual aspect of a complete solution of the so-called Einstein-Jordan
conundrum. It is well-known that Einstein steadfastly {\normalsize rejected
Born's assignment of probability to individual events, but it is difficult to
imagine that he would have refused a probability as an intrinsic property of
ensembles in a KMS state, in particular if it would have been clear to him
that this is an unavoidable consequence of the quantum adaptation of Faraday's
and Maxwell's "action at the neighborhood principle" and his own achievement
of its Minkowski spacetime causality reformulation. }

In fact vacuum polarization at the causal boundary, leading to the
aforementioned $\varepsilon\rightarrow0$ divergence of the dimensionless area
factor in the localization-caused entropy, as well as to the thermal
manifestation are \textit{two sides of the same coin}. In Jordan's model of
QFT the "localization-thermality" is not only an analogy to Einstein's
statistical mechanics calculation for black body radiation, but it is actually
isomorphic to a global heat bath thermal situation; i.e. in this special
situation there is a kind of inverse Unruh picture which permits to construct
a global heat bath system (Einstein's side) to Jordan's localization-caused
fluctuations in an interval on a lightlike line associated within his chiral
model (which he referred to as the wave quantization of a two-dimensional
Maxwell field).

A glance at a recent review of the early work on QFT \cite{Du-Ja} reveals that
even nowadays many authors firmly believe that the quantum mechanical result
that a global vacuum passes to a local vacuum (the vacuum of QM factorizes)
continues to be true in QFT and that only by coupling the QFT to an external
heat bath (applying the rules of statistical mechanics) can the
Einstein-Jordan conundrum be fully understood. This incorrect belief is
supported by referring to QFT as "relativistic QM" (with possibly infinite
degrees of freedom), a terminology which unfortunately has entered many
articles and even textbooks on QFT. But infinite degrees of freedom as a
result of the second quantization formalism applied to QM leading to
Schr\"{o}dinger quantum fields do not change the physical content, and the
correct distinction based on fundamental differences about localization is
considerable more subtle.

It is an important question to ask why such a foundational aspect of QFT was
only understood several decades after it first became visible in particular
situations as the Unruh Gedankenexperiment and the issue of quantum matter
behind (or in front of) black hole event horizons. The reason has to do with
the formulation of renormalized perturbation theory. Whereas the pre-Tomonaga
perturbation formalism (which was modeled on the quantum mechanical oscillator
formalism) failed whenever vacuum polarization properties entered the higher
order perturbative calculation, the covariant formulation, combined with
recipes involving ultraviolet cutoffs or regularizations and their final
removal through renormalization prescriptions in the works of Tomonaga,
Feynman, Schwinger and Dyson, shaped the form of renormalized perturbation.
Its first triumph in quantum electrodynamics and subsequently in the more
general gauge theoretic formulation of the standard model have made QFT the
most successful albeit unfinished theories of particle physics.

A detailed understanding of the derivation of these perturbative rules in
terms of the underlying causal localization principle (which could have led to
a better understanding of the consequences of causal localization) was not
really necessary; the Epstein-Glaser iterative implementation of this
principle ("causal perturbation") has remained little known; renormalized
perturbation does not require a foundational understanding of the causal
localization principle of QFT, a working knowledge of renormalized
perturbation theory in terms of calculational recipes suffices.

The situation changes radically if one comes to problems for which the
covariance of the formalism is of not much help, as the Einstein-Jordan
conundrum which addresses fluctuations from subvolume localizations. Even for
the simplest of all theories, namely that generated by a free scalar covariant
field, this is anything but simple. Since it cannot be exactly solved, it is
necessary to secure that the approximation is in agreement with the "holistic"
aspects of causal locality \cite{Ho-Wa}\cite{E-I} in a more direct way,
covariance alone does not help. Such aspects are easily overlooked in unguided
quantum mechanical calculations based on global oscillator degrees of freedom
as in \cite{Du-Ja}. This problem has its actual counterpart in the
impossibility to understand the cosmological constant by occupying global
energy levels of particles and enforcing the finiteness of the result by a
Planck length cutoff. \ \ 

The use of the new covariance property was sufficient to liberate the old
quantum mechanical based perturbation theory from problems caused by vacuum
polarization\footnote{In old textbooks of QFT (Heitler, Wenzel) the limitation
of the old quantum mechanical inspired formalism is visible in its restriction
to terms without vacuum polarization contribution (tree graphs).}, but
misunderstandings about causal quantum localization continued in areas of
S-matrix-based particle theories, notably the dual model and string theory. It
is the main concern of this paper to explain this in detail. Whereas the
understanding of these central issue in QFT after its birth in the aftermath
of the E-J conundrum was for a long time incomplete, it needed the appearance
of the dual model and ST to arrive at it genuine misunderstandings with grave
consequences for particle theory (of which only a science outside the natural
sciences as mathematics could profit\footnote{One may also add the
entertainment industry, whose link to natural sciences has been revolutionized
in prgrams directed by Brian Green.}).

Before getting to this issue, we have to recall some basic facts about why
particle physicists at the end of the 50s became increasingly interested in
attempts to access particle theory through a more direct use of the S-matrix
and on-shell approximation methods.

The rather limited, but within its self-proclaimed aims very successful
project, resulting from the particle physics adaptation of the
(Kramers-Kronig) dispersion theory, led to a revival of older (by that time
already abandoned) ideas to formulate particle theory directly in terms of its
most important experimentally accessible object: the scattering matrix
$S_{scat}.~$All post dispersion theory S-matrix attempts were based on an
important new property which had arisen from the setting of dispersion theory
as well as from Feynman's perturbation theory: the use of analytic properties
of scattering amplitudes, in particular the analytic continuation of the
\textit{crossing identity}. The first such scheme, the S-matrix bootstrap, was
already given up after one decade. The reason was not that any of its basic
assumptions turned out to be incorrect, but rather because they resisted a
coherent operational formulation which could be the starting point of
controlled approximations. Another reason was the strong return of gauge
theories, in particular in connection with the property of asymptotic freedom
for strong interactions.

For some years S-matrix ideas continued to be of phenomenological interest in
connection with a conjectured Regge behavior in certain regimes of scattering
amplitudes of strong interactions. \ However what brought these ideas to an
almost 5 decades lasting dominance of string theory was not phenomenology, but
rather a mathematical observation by Veneziano that by a clever use of
properties of gamma functions it is possible to construct a meromorphic
crossing symmetric function of two variables with infinitely many poles in
each variable interpreted as particles in an approximation of a new S-matrix
theory which was expected to result from an ill-defined unitarization of that
"would-be" one-particle approximation called dual model. Apart from the
presence of infinitely many pole-terms and the absence of threshold cuts
(whose presence turn out to be important for the meanwhile understood true
particle physics crossing), this was what Mandelstam expected in his setting
of two-variable spectral representations for elastic scattering amplitudes
\cite{Mandel}. In any case, the sanctioning by Mandelstam was very important
for the increasing popularity of this point of view.

This, at that time quite impressive mathematical construct of a \textit{dual
model,} was interpreted as a lowest order crossing symmetric solution which
still has to be subjected to an iterative "unitarization"\footnote{In contrast
to the derived Jost-Lehmann representation for matrix-elements of products of
two fields which was the rigorous basis of the derivation of the dispersion
relations, Mandelstam's spectral representation was assumed.}. This happened
at a time when the conceptional origin of the true particle crossing for $S$
and particle formfactors was not yet sufficiently understood. It was seen as a
concretization of spectral representations for the elastic scattering
amplitude, a project which Mandelstam had previously formulated in an attempt
to find a more specific setting beyond the generalities of the bootstrap
setting. Its name "dual model" referred to this incorrectly identified
crossing property unfortunately with particle crossing in a new (alledgedly
unique) S-matrix setting. The uniqueness was vetoed by several subsequently
found dual models, but reinstituted by the requirement of string theory which
added the requirement that the "mass spectrum" should come from a positive
energy representation of the Poincar\'{e} group on the oscillator degrees of
freedom contained in the Polyakov action (the square of the Nambu-Goto
action); This then led to the unique ten dimensional superstring representation.\ 

The dual model and its extension into string theory appeared much too
sophisticated for its phenomenological use in scattering theory of strong
interactions. Discrepancies with experimental scattering data and a return of
Yang-Mills gauge theories in connection with strong interactions led nearly to
its abandonment before it was elevated to a foundational $S$-matrix theory of
particle physics. The important point which led to this promotion was the
presence of an infinite particle spectrum which, in addition to most of the
observed particles also contained zero mass s=2 "graviton" together with the
promise of an ultraviolet converging approximate description of an
Einstein-Hilbert like quantum interaction. Never before in the history of
physics any model has been subjected to such a big interpretive jump over so
many orders of magnitudes as that of the change of string tension for strong
interacting "Regge strings" to the string tension required to describe Planck
length physics in a hypothetical quantum gravity.

The social success of such a gigantic jump does not depend on cohesive
physical arguments but on the size of the community willing to accept it and
the reputation of some of its more charismatic supporters. It was clear that
once accepted by part of the particle physics community, this would be the
first theory which was under total protection umbrella against observational
annoyances. Busting the conceptually tight corset of QFT, ST and its
derivatives ("target" embedding of lower dimensional source theories into
higher dimensional QFTs, the use of extra dimensions and the claim that
Kaluza-Klein dimensional reductions of classical theories commutes with
quantization) became a popular area of research in which community protection
permitted to ignore restrictions imposed by the successful principles of
particle theory.

To a large degree its popularity draws on its commitment to classical
mathematical-geometric ideas which are not burdened with the subtle
epiphenomena of quantum causal localization as vacuum polarization on its
boundaries and thermal manifestations of the localized vacuum state (e.g.
topological Lagrangians as the WZW action). In contrast to QFT, which had its
problems with interactions involving higher spins and only presently is in the
process of finding ways out by using the possibilities of short-distance
reducing properties of semiinfinite string-localized fields offered by modular
localization, ST became the candidate for the millennium "theory of
everything" (TOE).

Although there are by now increasing doubts even within the ST community about
whether this enormous popularity was scientifically justified (so that a
critical review like this may only support an anyhow ongoing critical trend),
it is not the purpose of this article to join the increasing community of
individuals who criticize string theory for its lack of success. Rather the
main aim is to expose the conceptual contradictions string theory always had
with the local quantum physical principles of particle theory and which in
normal times\footnote{"Normal" are times in which the ongoing computations are
on par with their conceptual understanding. "Shut up and calculate" on the
other hand is characteristic of the Zeitgeist of ST. Without ST, public
relation and entertainment activities on metaphoric subjects as extra
dimensions (Lisa Randall) and parallel universes (Brian Green) would not not
have appeared.} would have prevented its ascent. The only potential advantage
of its fading popularity in the present context is the small chance that its
proponents may be more open to foundational physical critique (instead of
propagandistic mudslinging).

Particle theorists often abandon a theory which, after many years of
exploration, failed to make contact with laboratory experiments or
astrophysical observations. This is not helpful for those who are left behind
and who would like to know whether the defeatist stance of their colleagues
was mainly resulting from their impatience with a deep theoretical, but
physically less successful project, or whether there were deeper more
foundational reasons within its conceptual structure which, although difficult
to formalize, activated their physical gut feelings telling them that it is
time to leave. There is of course the additional personal problem of
abandoning part or all of one's own personal history; for people who dedicated
more than 3 decades of their scientific life exclusively to one project, one
seems to be asking too much. A theory which had survived for such a long time
can (and may be should) not really disappear without a trace, at least
historians and philosophers of science will be curious to understand what
happened during all that time and what finally contributed to its disappearance.

Since the reasons for the loss of popularity are, unlike the abandonment of
the S-matrix bootstrap, not related to the emergence of a successful new
theory, a critical foundational review which unearths intrinsic fault lines
may even point into new research directions of particle theory.

ST fails on all counts of \textit{what its supporters claim it is}. What it
really represents, namely a special collection of infinitely many oscillators
associated to the canonical quantization of the Polyakov action which carries
a 10-dimensional positive energy representation of the Poincar\'{e} group that
is generated by an infinite component point-localized wave function (resulting
in an infinite component second quantized pointlike field), is not what can be
subsumed under the heading of ST. What its protagonists wanted to see, namely
spacetime strings in form of a chiral QFT on a lightlike line (or its circular
compactification) as a "source theory" embedded in its own "target space"
(alias inner symmetry space), contradicts the quantum principle of causal localization.

One may find the result that the oscillators in the supersymmetric extended
Polyakov action can carry a (unique, highly reducible) positive energy
representation of the Poincar\'{e} group in precisely 10 dimensions highly
remarkable, but unlike ST, the peculiarity should be explained in terms of
special properties of certain (non-rational) chiral models\footnote{Even the
most hardened reductionist would probably shun away from inverting his
expectation that a foundational theory should be rather unique in the sense of
permitting no similar theories in its neighborhood. \ But it is precisely this
inversion (rare, unique --%
$>$%
fundational) which leads string theorists to believe that we are living on a
dimensionally reduced 10-dimensional target space of a conformal QFT.}. It is
indeed hard to believe that ST could have attained its popularity on this dry
fact which even contradicts its name. Contrary to particle physicist who
reject ST on their physical instincts, the author draws his fascination with
this subject from the fact that ST fails because the birth-defect of QFT,
which was related to the incomplete understanding of the Einstein-Jordan
conundrum and which left renormalized perturbation theory largely unscathed,
finally led to deep misunderstandings about the dual model and ST.

As often in the history of physics (ether theory,...) the resolution of deep
misunderstandings is the source of rapid progress; in section 3 and at the end
of the paper there will be some indications that the correct understood string
localization of quantum fields (not embedding of QFTs) maybe the way out of
the present stagnation of gauge theory. As it is well-known pointlike quantum
fields of higher spin (s$\geq1$)~cause problems with renormalizability. At the
bottom of this problem is a clash between the Hilbert space structure
(positivity) and causal localization. Lagrangian quantization resolves this
clash by working in Krein spaces (Gupta-Bleuler, BRST) and returning to
physical pointlike localized subsystems by enforcing BRST gauge invariance at
the end of the calculation. Using string-localized potentials in an extended
form of causal perturbation theory, one stays within the Hilbert space setting
by giving up unreasonable restrictions on localization which come from
Lagrangian/functional quantization but have nothing to do with an intrinsic
understanding of QFT.

It may be interesting to the reader to familiarize himself with another point
of view which has nothing to do with strings but collects in a nutshell that
what can be rescued from a mathematical observation made by string theorists.
This is the solution of an old problem posed by Majorana \cite{Maj}: find an
irreducible algebraic structure which carries a (reducible) positive energy
representation of the Poincar\'{e} group whose decomposition leads to an
interesting one-particle spectrum. He was obviously thinking of th O(4,2)
spectrum of the hydrogen atom. This topic was taken up again in the beginnings
of the 60s by Barut, Kleinert, Fronsdal,..\cite{To}. The Ansatz in terms of a
noncompact group algebra which extends the Lorentz group failed, and the whole
project was abandoned. The irreducible system of oscillators in the Polyakov
action which leads to the superstring one-particle positive energy
representation of the Poincare group is the only known unique (up to a finite
number of M-theoretic variations) solution. It is easy to see that one needs a
chiral model with a continuous superselected charge spectrum in order to
encounter noncompact group representation as the Poincare group acting on its
inner symmetry ("target") space; multicomponent abelien current algebras are
the only known models which fulfill these requirements. The former interest in
this old project of finding "dynamic"\footnote{Referring to the requirement
that the infinite (m,s) particle tower should arise naturally from a larger
indecomposable algebraic structure.} infinite component field equations is
nowadays difficult to convey since these ideas originated at a time long
before the subtleties of the particle-field relation in the presence of
interactions were appreciated.

Sometimes one can spot the origin of misunderstandings without much effort.
When the Lagrangian of a classical relativistic particle $\sqrt{ds^{2}}$(see
next section) is presented as a "warm up" for a covariant quantum string
theory \cite{Polch} which was expected to result from the quantization of a
Nambu-Goto action, the conceptual alarm bell should have started ring. It was
one of the great achievements of Wigner to understand that relativistic
particles cannot be described by quantization of a classical covariant
particle action, but one rather has to resort to representation theory of the
Poincar\'{e} group and even that only works in the absence of interactions.
The only covariant description which has (asymptotically attached) interacting
particles is interacting QFT. Hence why should one believe that covariant
string-localized analogs of particles can be obtained from a quantization of
the Nambu-Goto action?

To present the problem the other way around: there simply do not exist
covariant 4-component covariant operators $x_{op}^{\mu}$ whose spatial
components are quantum mechanical position operators. This is well-known and
easy to verify by looking at the spectral projections $E$ from the spectral
theory of selfadjoint operators%
\begin{align}
&  \vec{x}_{op}=\int\vec{x}dE_{\vec{x}},\text{ }R\subset\mathbb{R}%
^{3}\rightarrow E(R)\\
&  U(a)E(R)U(a)^{-1}=E(R+a),~E(R)E(R^{\prime})=0\text{ }for~R\times R^{\prime
}\nonumber\\
&  \left(  E(R)\psi,U(a)E(R)\psi\right)  =\left(  \psi,E(R)E(R+a)U(a)\psi
\right)  =0\nonumber
\end{align}
where the second line expresses translational covariance and orthogonality of
projections for spacelike separated regions. In the third line we assumed that
the translation a shifted $E(R)$ spacelike to itself. But since $U(a)\psi$ is
analytic in $\mathbb{R}^{4}+iV^{+}~$($V^{+}~$forward light cone) as a result
of the spectrum condition, $\left\Vert E(R)\psi\right\Vert ^{2}=0$ for all $R$
and $\psi~$which implies $E(R)\equiv0$ i.e. covariant position operators do
not exist. Hence this analogy is counterproductive for strings based on the
quantized Nambu-Goto action. What the correct quantization of the latter
really gives has nothing to do with what string theorist expect nor is it a
dynamic infinite component solution of the Majorana problem (next section).

QFT was born in the aftermath of the Einstein-Jordan conundrum \cite{E-J} with
an insufficient awareness about the consequences of causal
localization\footnote{This dispute led Jordan to the dicovery of QFT (at thet
time matter-wave quantization) \cite{Schwe} \cite{Du-Ja} \cite{Jo}.}. The
reason why this is relevant in connection with string theory is that whereas
the incomplete understanding of the deeper layers of causal-localized QT did
not really cause problems with the recipes of renormalized perturbation
theory, this is the first time that it had serious consequences. It required
the appearance of the dual model and ST and its widespread uncritical
acceptance to cause a real derailment in particle theory. There remains of
course always the hope that even at this late hour the comprehension and
correction of deep misunderstandings may lead to new hints for future
directions in particle theory.

Since the correct formulation and conceptual understanding of S-matrix-based
ideas is pivotal for the present work, a substantial part will be dedicated to
the presentation of \textit{particle crossing} which, besides scattering
theory, is the foundational link between particles and fields (as the
generators of local observables). The crossing identity of particle theory and
its analytic continuation is one the deepest and subtlest relations between
particles and fields. Its derivation from causal localization requires an
intrinsic understanding of the latter. This has been achieved in the local
quantum physical (LQP) setting of QFT in the form of modular localization as a
part of modular theory of operator algebras (next two sections). It is an
important property in any new constructive approach to QFT which replaces the
functorial relation between Wigner's intrinsic\footnote{"Intrinsic" in the
present setting means: independent ot the classical parallelism on which
Lagrangian quantization is based.} representation theory of particles and free
fields its functorial relation with free fields by an S-matrix based
construction of generators of wedge-localized algebras which results from the
fact that the S-matrix is a relative modular invariant of wedge-localization.

Without such a conceptual investment it would be hard to understand at what
point Mandelstam's courageous project to attribute a constructive role to the
S-matrix (leaving aside all references to Lagrangian/functional quantization
and other parallelisms to classical physics) eventually failed; in fact some
of the aspects of the new "top-to-bottom" approach in section 3 may be
interpreted as a resurrection and extension of those old on-shell ideas
(S-matrix, formfactors) which in the 60s served as an antidote against the (at
that time perceived) threat of the ultraviolet catastrophe in off-shell QFT in
terms of correlation functions of fields.

Whereas his idea to use spectral representations \cite{Mandel} for scattering
amplitudes as a means to control their analytic properties was still well
within the spirit of the time, his embracing of Veneziano's mathematical
guesswork \cite{Ven} on crossing trapped him into a wrong type of crossing
which, as we know nowadays, has nothing to do with particle theory. It rather
describes the crossing property of Mellin transforms of conformal QFTs, i.e.
of theories of scale invariant anomalous dimension QFT which (apart from
conformal free fields) do not describe particles (next section). Insufficient
understanding of the quantum aspects of causal localization for particle
crossing properties did not leave much of a chance to get out of this
conceptual trap at that time.

S-matrix-based settings have a long and eventful history, but that of LQP, an
intrinsic formulation of QFT which shares with Mandelstam's S-matrix setting
the abstention from quantization analogies with classical field theories, is
even longer. It started with Haag's 1957 attempt\footnote{The original version
is in French even though most of the talks were in English. Later it was
translated back into English in \cite{1957}..} \cite{Freden} to base QFT on
\textit{intrinsic principles} instead of linking a more fundamental theory via
a \textit{quantization parallelism} to a less fundamental one. Hence the
terminology LQP stands for a \textit{different formulation} of QFT
\textit{which maintains its physical content}. Another setting which also did
not refer to quantization was Wightman's \cite{St-Wi} formulation of quantum
fields in terms of operator-valued Schwartz distributions and their
correlation functions.

Both ideas, Mandelstam's as well as Haag's (with important early contributions
by Borchers \cite{Haag}), were top to bottom approaches in the sense that one
states initially those properties which could be helpful for particle theory
before looking for computational tools to implement them. The main difference
was that Mandelstam's proposal was based on the S-matrix, which from the
viewpoint of local quantum physics is the \textit{observational crown of a
QFT} (and not its foundation).

Haag's local quantum physics (LQP) approach was modeled on the enormous
successful \textit{action at the neighborhood} principle of Faraday and
Maxwell which he reformulated as a \textit{principle of causal localization of
quantum matter}. Its formulation required a lot of modern (in parts unknown at
that time) mathematics. Different to Mandelstam, LQP had no use of the
S-matrix as a computational tool. Often the intuitive content of an idea is
partially lost, while its mathematical formulation became increasingly
precise; this seems to be the fate of all foundational concepts, and the idea
of \textit{modular localization,} which leads to most of the results in this
article, is no exception. It is precisely this idea which leads to a
completely new constructive role of the S-matrix.

In this paper it will be shown that there is indeed a quite unexpected
synthesis of the two views. It is based on the recognition that, besides its
prominence in the large timelike asymptotic behavior of scattering theory, the
S-matrix is also a "relative modular invariant of wedge localization"
\cite{S1}. This leads to new nonperturbative dynamical ideas in LQP into
which$~S_{scat}$ enters on par with other foundational properties \cite{S2}. A
first success of this new setting is the \textit{existence proof} of a
particular family of two-dimensional models (factorizing models) with a
realistic (noncanonical) short distance behavior \cite{Lech}. \ 

Both attempts, Haag's and Mandelstam's, avoid the classical-quantum
parallelism of Lagrangian quantization which constitutes the basis for the
standard perturbation theory; in Lagrangian quantization one starts from the
Lagrangian formulation of classical field theory {\small {\normalsize and
explores what one gets by following the quantization rules and imposing
reasonable interpretations on the computational results. The same renormalized
perturbation theory (but without intermediate cutoffs or regularizations) can
be derived from an iterative implementation of causal locality (the
Epstein-Glaser approach \cite{E-G}) starting from a Wick-ordered polynomial in
pointlike free fields which specifies the form of the interaction. An
on-mass-shell perturbative calculation of the S-matrix only (without
introducing other on-shell objects) is not possible; only after having
computed correlations one can study their restrictions to mass shells. }}

My own contribution \cite{S1}\cite{S2} consisted in trying to bring these two
nonperturbative ideas together. As mentioned before, this led in the hands of
Lechner \cite{Lech} to an existence proof for certain d=1+1 models with
noncanonical short distance behavior. In this way the extensive work on the
bootstrap-formfactor project, in which formfactors of d=1+1 factorizing models
were successfully calculated on the basis of recipes (for recent review see
\cite{Ka}), received a solid conceptual-mathematical underpinning. Another
result of this intrinsic way of dealing with QFT is a foundational
understanding of the issue of integrability \cite{integrable} and ideas about
how to approach nonintegrable theories (which includes all observational
relevant models in particle physics).

There is a well-founded hope that an existence proof and a controlled scheme
of approximations for the general case may come out of these
attempts\footnote{Neither in case of classical nor for quantum mechanical
nonintegrable systems one can hope for more.}. Far from being a chill pill of
finnicking axiomatists, there is good reason to expect that such a step will
finally lead to such a conceptual closure of QFT, a role which unfortunately
the divergent series of renormalized perturbation theory cannot play.

The incorrect ideas which led to the dual model and string theory took the
form of three different but interlinked proposals. The first one is the dual
model. There are many dual models; for each conformal QFT there is one,
independent of its space time dimension \cite{Mack}. Their definition in terms
of Mellin transforms shows that the pole spectrum of the meromorphic functions
is given by the anomalous dimensions of (composite) operators appearing in
convergent global operator expansions of products of conformal fields;
\textit{it has nothing to do with the particle spectrum} of an approximand of
the two-particle scattering amplitude. The relation of Mellin transforms to
states and Hilbert spaces is totally different from that of scattering
amplitudes. The representation theory of the Poincar\'{e} group plays no role
in this construction of meromorphic functions.

The second idea is based on the canonical quantization of the Nambu-Goto
Lagrangian resulting from ignoring the square root. This approach is more
ambitious since it does not only aim at the crossing in terms of poles of
particle masses but also at a unitary positive energy representation of the
Poincar\'{e} group related to the pole positions. This led to a system of
oscillators (apart from a quantum mechanical p,q zero mode) on which one may
represent a highly reducible unitary positive energy representation of the
Poincar\'{e} group in a 10-dimensional spacetime: the famous
\textit{superstring representation }(unique up to a finite number of
M-theoretic modifications). The space generated by the oscillators contains in
addition to the degree if freedoms which are used to built up the generators
of the Poincar\'{e} group also operators which connect \textit{the different
levels} of the infinite mass/spin tower. The field which arises from second
quantization of the wave function space is a \textit{pointlike-localized
dynamical infinite component field}. By projecting onto finite mass subspaces
it becomes a finite component Wightman field (operator-valued Schwartz distribution).

The third construction gives a maximum of insight. It consists in starting
from a non-rational (continuously many superselection rules) chiral theory.
The concrete model which was used for this purpose is the model of an
n-component abelian chiral current. Defining a \textit{sigma-model field}
through the exponentials of the potentials of the currents, one then may ask
whether it is possible to have noncompact inner symmetries acting on the index
space of the potentials. Surprisingly the answer is indeed positive; it is
perfectly possible to represent noncompact inner symmetries acting on the
index space of such non-rational chiral theories (referred to as "target
space" by string theorists). In fact one can even represent the Poincar\'{e}
group; and the only real surprise is that on the multi-component chiral
current theory there exists a positive energy representation of the
Poincar\'{e} only for n=10, and it is rather unique (up to a finite number of
"M-theoretic" modifications). This construction has the closest relation to
the modular localization of the next section. This quantum mechanical
representation of the Poincar\'{e} group does not contain representations with
"infinite spin" which are the only Wigner representation which upon second
quantization would lead to semiinfinite spacelike string-localized
representations. Hence this representation does not contain stringlike
localized components and hence is point-like generated.

This third path to string theory is most revealing because it shows that the
quantum mechanical positive energy representation theory of the Poincar\'{e}
group handles the zero mode of the Fourier decomposition of the chiral current
theory in a different way as required by the pointlike nature of chiral model.
In other words the pointlike nature in "source" space is not compatible with
this representation of the Poincar\'{e} group in target space. This is one of
the reasons why there is no embedding of the chiral theory in the target
theory in a literal sense. A deeper reasoning shows that the holistic nature
of localization in QFT \cite{Ho-Wa}, in contrast to the Born localization of
quantum mechanics, \textit{never allows} embeddings of lower dimensional into
higher dimensional theories; \textit{neither is a Kaluza-Klein reduction of
dimensions consistent with the holistic quantum structure of causal
localization} although it is perfectly consistent with classical field theory
and the Born localization of QM (and even with quasiclassical approximations
of QFT). The embedding of a linear quantum mechanical chain of oscillators
into a space of arbitrary dimension is only possible in QM where
"localization" lacks an intrinsic meaning.

Before presenting the arguments against ST, it way be helpful to state the
conclusions in the form of a collection of theses which will receive detailed
attention in the subsequent sections

\begin{itemize}
\item The crossing symmetry of the dual model amplitude has no relation to the
crossing identity in particle physics. In fact what is interpreted as a (m,s)
"particle-spin tower" is really the tower of a $(d_{sd},s)$ (anomalous) scale
dimension spectrum which occurs in the global operator expansion of two fields
in a conformal QFT \cite{Mack}. ST shares a particular version of this spectrum.

\item Different from what its name suggests, string theory does not describe
string-localized objects in spacetime. In particular an embedding of a chiral
conformal QFT into a higher dimensional target space is incompatible with the
formulation of causal localization in QT.

\item ST is a dynamic infinite component QFT, in fact it is the only known
solution of an old problem which goes back to Majorana \cite{Maj}, namely a to
find an irreducible algebraic structure which carries a representation of the
Poincar\'{e} group with an interesting (m,s) spectrum.

\item The quantization of the diffeomorphism-invariant Nambu-Goto action (the
original action for ST) has no relation to conformal chiral theories and leads
to very different results from those obtained from the Polyakov action; in
particular there is no discrete (m.s) spectrum.

\item The Maldacena conjecture (originally a derivation of ST) as a
correspondence between two \textit{physical} theories is contradicted by known
facts about the connection between the cardinality of phase space degrees of
freedom and the "causal completion property" (timelike causality). Either the
conformal QFT has too many degrees of freedom (overpopulation leading to
causal "poltergeists") or the upload of a normal conformal QFT leads to a AdS
theory which is too "anemic" to support causal localization as we know it
(last section).
\end{itemize}

The incomplete understanding of epiphenomena of modular localization has also
led to misunderstandings in derivatives of ST, as Maldacena's version of the
AdS-CFT correspondence, "brane physics" and extensions of the classical
Kaluza-Klein dimensional reduction idea to quantum theories based on causal
localization\footnote{The reason why we occasionally use this round-about
terminology is that the reader may be unaware that our understanding of QFT
extends significantly beyond what can be found in textbooks..}. Although these
concepts are only historically and sociologically but not logically tied to
ST, their critical evaluation follows similar conceptual reasoning and will be
presented after having settled the above claims.

\section{Crossing from conformal correlations and crossing in particle
physics}

One of the manifestations of relativistic QT which is directly related to the
foundational is the particle crossing property; it requires to understand the
relation between particles and fields beyond their asymptotic relation in
terms of scattering theory.

This property, was first observed on Feynman graphs: graphs with a certain
number of external lines restricted to the mass-shell describing a cluster of
incoming and outgoing particles, can be considered as a relation between a
formfactor of an operator which is associated with the remaining unrestricted
(off-shell) lines. Shifting some lines from the incoming to the outgoing
particle configuration formally corresponds to a process where a lesser number
of particles is incoming and a larger outgoing, except that the shifted
outging objects are on the backward mass shell and (this is the nontrivial
part of "crossing") can be related to the crossed physical formfactor by
analytic continuation. The crossing from outgoing to incoming particles is
analogous. The S-matrix is the formfactor of the identity operator; in case of
the elastic S-matrix the energy-momentum conservation requires to cross a pair
of particles, one from the incoming and the other from the outgoing
configuration. The formal graphical crossing is trivial; the nontrivial part
is the possibility of an analytic continuation \textit{inside} \textit{the
complex mass shell;} this is nontrivial even in perturbation theory.

In the context of the present paper the main interest is not the result as
such (which seems to be "obvious" in a graphical presentation), but rather the
subtle concepts used in its proof, as well as the nontrivial
\textit{thresholds modifications }in case that the particle wave functions
overlap. The new concept of \textit{modular wedge localization} leads to a KMS
identity in terms of fields, whose appropriate transcription into particle
formfactors reveals the true conceptual origin of particle
crossing\footnote{For the special case of the elastic S-matrix there exists a
proof based on the theory of analytic functions of several variables
\cite{BEG}. Probably as a consequence of its somewhat unbalanced relation
between mathematical expenditure and the limitation of its physical range,
this rigorous proof remained largely ignored by practinioners of QFT.}. The
ideas are intimately related to a foundational understanding of the
formfactor-bootstrap constructions for d=1+1 factorizing models \cite{BFKZ}%
\cite{Lech}. In the wider context of integrable and nonintegrable QFTs
\cite{integrable}, their presentation will be deferred to the next section.

Whereas the old arguments for crossing in the context of the bootstrap
S-matrix project remained vague, the situation changed when this property was
used in constructions of "would-be" scattering amplitudes. In the proposed
\textit{dual model} \cite{Ven} it became sufficiently concrete, so that with
the hindsight of the present day knowledge it is fairly easy to see that the
dual model crossing has no connection with the particle crossing as first seen
in Feynman diagrams and later used as a basic property in the formulation of
the S-matrix bootstrap. To see that it has no relation with particle crossing
does not require as much preparation as the proof of the real crossing
property, and therefore most of the critique of the Dual Model and ST will be
presented already in this section.

Even though the foundational origin of particle crossing was not understood at
the time of the dual model, it was clear that it describes a subtle interplay
between one particle poles of the S-matrix or more general between formfactors
which cannot be approximated by a meromorphic function. Veneziano's
meromorphic dual model was the result of an educated mathematical experiment
guided by the properties of the Euler beta function. Although the result of
this experimental mathematics does not fit what one expects from particle
crossing, but one may still ask whether there is anything else in local
quantum physics which this model describes. The answer is surprising as well
as interesting; the dual model encodes a property of conformal QFT, an area of
local quantum physics which has no (known) relation to particle theory.

Schematically this may be described as follows \cite{Mack}. Quantum fields in
conformal QFT have different properties from those which admit a particle
interpretation. As a result the only asymptotic converging Wilson-Zimmermann
short distance expansions is replaced by a \textit{globally converging}
operator expansion in which the coefficient functions have the properties of
conformal 3-point functions%
\begin{align}
&  A_{3}(x_{3})A_{4}(x_{4})\Omega=\sum_{k}\int d^{4}z\Delta_{A_{3}%
,A_{4.},C_{k}}(x_{1},x_{2},y)C_{k}(z)\Omega\\
&  \left\langle A_{1}(x_{1})A_{2}(x_{2})A_{3}(x_{3})A_{4}(x_{4})\right\rangle
\rightarrow3\text{ }different\text{ }expansions\nonumber
\end{align}
where the second line expresses the fact that, by either using local spacelike
commutativity of fields in Minkowski spactime or commutativity of their
euclidean counterparts, one obtains a situation which resembles the three s,
t, u Mandelstam variables in the parametrization of the s,t,u physical
scattering channels related by analytic continuation. The analogy becomes even
closer upon Mellin transforming in the above spacetime variable, in which case
the numerical values of the scale dimensions of the composites in the global
expansion pass into positions of first order poles; the result consist in
three different converging pole expansions of the same meromorphic function
into sums over pole contributions. Unlike Fourier transforms, Mellin
transforms have no independent operator status in Hilbert space since they are
performed on correlation functions and not on individual operators (i.e. there
are no "Mellin-operators" in the sense that there exist Fourier transforms and
momentum space operators).

This implies in particular that there is no Hilbert space description which
can place Mellin-transforms of correlation functions and S-matrix amplitudes
under one conceptual roof (which would be the minimum requirement for a basis
of unitarization of the dual model amplitude). Independent of the
dimensionality of the conformal model, the so obtained functions are always
meromorphic with infinitely many first order poles at positions given by the
anomalous scale spectrum of those fields which appear in the global operator
product decomposition. For mathematical details about Mellin transformations
of conformal correlations see \cite{Mack}.

The crossing symmetry resulting from the combined result of the global
operator expansion with subsequent Mellin transformation may have the effect
of an alluring siren/mermaid song, suggesting particle crossing in the sense
of particle physics to the conceptually untrained particle physicist. Its
wide-spread acceptance showed that the time of the European "Streitkultur" in
its use for a conceptual cleansing in particle theory had defintely come to an
end \footnote{Besides names as Landau, Lehmann, Jost, Kallen, ..this
clarifying instrument of the particle.theory discourse also took some roots in
the US (Oppenheimer, Feynman, Schwinger,..). It ended with Jost's masterful
polemic article against the S-matrix bootstrap \cite{Jost}.}. Starting with
Veneziano's famous dual model paper, a new Zeitgeist in which the use of
analogies was not the exception but rather the rule took over; analogies which
were not interdicted on the mathematical side became acceptable.

It turned out that there are many more dual models than that found by
Veneziano; essentially any conformal QFT in any spacetime dimension led to a
meromorphic crossing symmetric functions. As mentioned in the previous section
an important additional restriction came from the requirement that the
anomalous dimension spectrum ($d_{sd},s$) of conformal composites should be
identical to the the $(m,s)$ spectrum of a unitary representation of the
Poincar\'{e} group. It turns out that the (physically unmotivated)
identification of the two requirements is extremely strong; in fact there is
no a priori reason why it should have any solution at all, since the two
requirements have no visible relation.

The next step which led from the dual model to ST was the search for a
Lagrangian whose canonical quantization supports such a situation. The
somewhat surprising answer (found by exploring properties of the Virasoro
algebra) was that the Polyakov action (the square of the Nambu-Goto action)%
\begin{equation}
\int d\sigma d\tau\sum_{\xi=\sigma,\tau}\partial_{\xi}X_{\mu}(\sigma
,\tau)g^{\mu\nu}\partial^{\xi}X_{\mu}(\sigma,\tau) \label{po}%
\end{equation}
for $\mu,\nu=1....26~$repectively its supersymmetric extension in $10$
dimensions does the job; in the latter case one is even in the fortunate
situation of a positive energy representation which means (since no infinite
spin component occurs in the decomposition into irreducibles) that the
localization of 10 dimensional wave function and the associated second
quantized free fields is automatically pointlike\footnote{The classical
solutions of the Nambu-Goto action coalesce with those of the Polyakov action;
but this does not mean that this associated QFTs (see later).}. For the
following we do not have to know what formal arguments led to a representation
of the Poincar\'{e} group associated with \ref{po} and how convincing it is;
it suffices to accept the result that the oscillator degrees of freedom of a
26 component chiral current or its supersymmetric 10 component extension does
the job.

Also the well-known infrared problems of d=1+1 massless fields (potentials of
well-defined currents) are irrelevant; the curents as well as their associated
charge-carrying sigma-model fields (see below) which feature in our arguments
are well-defined (using the appropriate limiting definition starting from
finite exponential strings and shifting one charge to infinity), there is no
infrared problem.

ST begun with the recognition that the analogy between the poles at certain
anomalous dimensions and "would be" particle masses can be strengthened by the
construction of a unitary positive energy representation on the oscillators
degrees of freedom contained in a (supersymmetrically extended) 10 component
chiral current model. Accepting the result of the arguments of string
theorists, still leaves two questions connected with such a construction:

\begin{enumerate}
\item What is the property which enables chiral current theories to allow
representations of noncompact groups on there internal symmetry ("target")
space in view of the fact that the inner symmetries of higher dimensional QFT
describing particles can only accommodate compact groups.

\item Both, the localization of a chiral theory on a lightray (or circle) and
the wave functions (and second quantized fields) associated with the pointlike
generated target space representation of the Poincar\'{e} group realize
holistic localization properties in very different spacetime contexts. Do they
really use the same oscillator substrate supplied by the Polyakov action?
\end{enumerate}

The only possible "noncompact indices" of a field localized in n-dimensional
spacetime $n>3$ are those which refer to their tensorial/spinorial nature in
that dimension. How is it possible that for chiral theories they can support
noncompact and even Poincar\'{e} group representations of higher dimensional
theories? Here the distinction between rational and nonrational theories come
into play. The characterizing property of rational chiral theories is that
their observable local algebras have only a finite number of superselection
sectors which are generated by fields with generally plektonic (braid group)
commutation structures. The superselected charge spectrum must be continuous
in order to accommodate noncompact symmetry groups and the only known QFT (in
any spacetime dimensions) which are able to provide such a situation are
certain non-rational chiral models, most prominently multi-component current
models
\begin{align}
\partial\Phi_{k}(x)  &  =j_{k}(x),~\Phi_{k}(x)=\int_{-\infty}^{x}%
j_{k}(x),~Q_{k}=\Phi_{k}(\infty)\,,\ \Psi(x,\vec{q})=e^{i\vec{q}\vec{\Phi}%
(x)}\label{ana}\\
Q_{k}  &  \sim P_{k},~dim(e^{i\vec{q}\vec{\Phi}(x)})=\vec{q}\cdot\vec{q}\sim
p_{\mu}p^{\mu},~(d_{sd},s)\sim(m,s)\nonumber
\end{align}
The second line indicates the analogy on which the construction of a
10-dimensional positive energy representation (the superstring representation)
of the Poincar\'{e} group is based.

Coming to the second question, the answer is the following. The representation
of the Poincar\'{e} group based on the use of the irreducible algebraic
structure from the Polyakov action is sufficiently different from the holistic
requirements needed to describe localization of a chiral theory in order to
prevent an isomorphism or even an embedding of the chiral theory into its
target spacetime (related to the positive energy representation of the
Poincar\'{e} group). This is obvious from the different spectra of the zero
mode operator; whereas the charge spectrum of the oscillator degrees used to
implement the holistic localization structure of the chiral current theory is
continuous, the energy-momentum spectrum in the target space construction is
that of the direct sum of one-particle representations contained in the
superstring representation of the Poincar\'{e} group. Nevertheless there is
one property which the target theory preserves of its chiral avatar: the
$(m,s)$ spectrum, alias the $(d_{sc},s)$ chiral spectrum.

So the remaining question is: can the "picture puzzle" resulting from a
$(d_{sd},s)\sim(m,s)$ kaleidoscope be the starting point of a new foundational
particle physics theory? For string theorists the answer obviously
affirmative, they are probably not even aware about its existence. For the
rest of the world a non-negative response could consist in the perception
that, although these analogies do not solve that what the string theorists had
hoped for, they do lead to the \textit{first and only illustration of an old
forgotten dream by Majorana} about a "dynamic infinite component field
project" \cite{To} namely to find an irreducible algebraic structure which
carries a reducible infinite component one-particle representation of the
Poincar\'{e} group. This at that time interesting project (in analogy to the
$O(4,2)$ dynamical group of the hydrogen spectrum) from an epoch which had a
more naive understanding of QFT (relativistic QM) has disappeared in the
maelstrom of time after enjoying a second wave of popularity in the hands of
Barut, Fronsdal, Kleinert,.. \cite{To} in the first half of the 60s. In any
case it could not have been used as a scientific support for the sexed up
colorful science fiction stories of Brian Green and Lisa Randall which appear
in TV programs and in interviews and for which the ST Zeitgeist will always be
remembered, independent of the scientific future fate of ST.

The second line in (\ref{ana}) contains the definition of the charge-carrying
sigma model fields which generate the charge sectors and also play the role of
the formal carriers of the Poincar\'{e} representation in the nonisomorphic
target space interpretation. In contrast to the massless potential $\Phi$ the
sigma model fields $\Psi$ are well-defined. As a result of the nature of their
relation to the current fields they are not local but rather lead to abelian
braid group representations. Instead of quantizing the Polyakov action one
could also start from the sigma model field in which case the previous
observations would amount to saying that the holistic organization of sigma
model degrees of freedom which is necessary to support the localization in the
wave function space of the Poincar\'{e} group representation. Such points are
easily overlooked if one naively interprets the similarity of formal
appearances as an identity of the conceptual content which is the root cause
which led to the incorrect embedding idea. As a result of the wrong embedding
picture the spacetime interpretation of the operators $X_{\mu}(\sigma)~$as
tracing out a stringlike localization is incorrect\footnote{Apart from the
nullmode which describes the Fourier component of point-localization the
"movement" of the other oscillator variables is in an internal space "over"
the localization point (that where one pictures spin components).} and
unfortunately this also damages to the formal Euler beta-function like
argument that ST implies QGr.

There remains the philosophical question of what to make out of the near
uniqueness of the resulting 10 dimensional superstring representation (more
correctly uniqueness modulo a finite number of so-called M-theoretic
modifications). If the number would have been zero or infinity the whole story
about ST would have ended right there. The metaphoric idea of string theorists
that we are living in a (suitably dimensionally reduced) 10 dimensional
"target space" of a chiral current theory has to be seen in the context of a
widespread belief that foundational theories are unique or nearly unique in
the sense that there are no other theories in the immediate neighborhood.
Whereas some among us would subscribe to the idea%

\begin{equation}
fundamental~theory~(TOE)\rightarrow rarity~of~realization
\end{equation}
not even the most hardened reductionist would probably accept the inversion of
this arrow. But this is precisely what keeps ST going. The wisdom of many
vernaculars must be called into questions in times of ST. One is certainly the
saying \textit{many people cannot err}. In other times people would have
looked for an explanation of the rareness as a peculiar property of
nonrational models (whose observable algebras have a continuous cardinality of
superselection sectors).

There remains the problem of what becomes of ST if one takes the problem of
quantization the Nambu-Goto action (the square root of the Polyakov action)
serious. This has been recently answered in a paper by Bahns, Rejzner and Zahn
\cite{BRZ}. In that case there is no relation to a conformal QFT as in the
case of the Polyakov interaction. The quantization problem is similar to that
of quantizing a nonpolynomial interaction as the Einstein-Hilbert action
\cite{Fred}; the overriding problem is whether the aspect of background
independence can be separated from the obvious nonrenormalizability of both
actions. In both cases there is the indication that this is indeed the case
i.e. that diffeomorphism invariance can be implemented without having any
restricting effect on the renormalizability problem. Unlike the ST based on
the Polyakov Lagrangian, which at least solves the Majorana problem of a rich
one particle spectrum from an irreducible algebraic structure, the Nambu-Goto
action solves the problem of what remains of a classical Kaluza-Klein
dimensional reduction interpreted as a classical diffeomorphism invariant
embedding if one subjects it to the rules of diffeomorphism invariant quantization.

Obviously quantization of a classical embedding is not the same as the
embedding of a ready made lower dimensional QFT into a larger one. The problem
which is at the bottom of this interdiction also seriously impedes its
inversion i.e. the direct Kaluza-Klein reduction of a QFT in terms of
operators or correlation functions (and not by massaging Lagrangians). As far
as the Poincar\'{e} group representation in the special case of embedding into
a Minkowski spacetime is concerned, the result of some (not completely
understood "zero mass stuff") is not encouraging for a ST believer\footnote{In
fact the ideology of little strings swirlingly through spacetime is so strong
that string theorist do not accept the result of their own correct
calculations of pointlike (graded) commutators of what they insist to call
string fields. Instead of presenting that point as the middle of an invisible
string \cite{Lowe}\cite{Martinec}, they should have written (not being able to
let loose on the terminology "string") "internal string" which lives there
where one spin components live.}.

The quantization of a parametrization-invariant classic embedding has a close
conceptual (but not mathematical) proximity to an old setting proposed by
Pohlmeyer which was based on the observation that the Nambu-Goto Lagrangian
describes a integrable classical system. Pohlmeyer \cite{Pohl} identified the
infinite number of conserved currents and determined their Poisson bracket
relations which he than quantized in the spirit of integrable model
quantization. He did not identify a concrete representation of the
Poincar\'{e} group, so that the problem of localization (which in the ST case
refers to a concrete positive energy representation) of states remained open.

One would think that finding a physical structure which \textit{precisely}
leads to the dual model (and not to an imagined elastic approximation of a
crossing symmetric unitary S-matrix) as the global operator expansion in
conformal QFT \cite{Mack} that this settles the problem; but the interest in
conceptual aspects of local quantum physics obviously reached such a low point
("shut up and calculate" \cite{Teg}) that even this does not reach a community
which has throne its fortune to the side of mathematics. It seems that the ST
community which started with Veneziano's mathematical tinkering is dead-set to
also go down with it. How else, than in this way, can one explain that the
vague resemblance of multicomponent chiral currents and anomalous dimensions
with particle momenta and mass spectra of particles can trap so many people
(among them brilliant minds) if it is not the blind faith that mathematics
will fix it.

Within all the paeans of praise from mathematicians about how much they owe to
ST it may be helpful to point out that the correct use of the multi-current
model which is in harmony with its true role as a "theoretical laboratory of
QFT" is no way less sexy that its ST avatar. The extensions of the observable
algebra through the addition of sigma model fields with integer scale
dimensions are classified in terms of even lattices whereas their charge
structure is characterized in terms of the dual lattice\footnote{These
concepts were introduced for the one-component current in \cite{BMT} an
generalize to multicomponent currents in \cite{Sta}\cite{Ka-Lo}}. For quantum
field theorists it is very interesting to have selfdual examples because they
are fully "Haag-dual" (the only sector is the vacuum sector) which means that
not only for simply connected (interval) but also for multiply connected
(multi-interval) localized regions the commutant is equal to the algebra
localized on the complement. As far as I know these are the only Haag dual
algebras within the huge family of chiral models. The associated lattices and
symmetry groups correspond precisely to the largest semisimple finite groups.
This illustrates again the close connection of causal localization with group
theory as a special case of the DHR superselection theory which classifies
local representation equivalence classes of observable algebras in terms of
compact groups.

\section{Good news for higher spin interactions from modular localization}

Up to now we only used the critical potential of the modular localization
setting. Unfortunately ST cause a lot of confusion in an area in which
bona-fide string-localization really matters: interactions involving higher
spin fields $(m,s\geq1)$, in particular for gauge theories and their massive
counterparts as they are needed in the standard model.

The project to use the new setting of modular localization to solve remaining
problems of perturbative QFT started with the solution of a conceptual problem
which, since the days of Wigner's particle classification remained unsolved:
\textit{the causal localization of the third Wigner class} (the massless
infinite spin solutions) of positive energy representations of the
Poincar\'{e} group whereas the massive class as well as the zero-mass finite
helicity class are pointlike generated. Spacelike string-generated fields are
covariant fields $\Psi(x,e),$ $e~$spacelike unit vector which are localized
$x+\mathbb{R}_{+}e$ in the sense that the (graded) commutator vanishes if the
full semiinfinite strings (and not only their starting points $x)$ are
spacelike separated \cite{MSY}%
\begin{equation}
\left[  \Psi(x,e),\Phi(x^{\prime},e^{\prime})\right]  _{grad}=0,~x+Re~\rangle
\langle~x^{\prime}+Re^{\prime}%
\end{equation}
Unlike decomposable stringlike fields (line integrals over pointlike fields)
such \textit{elementary stringlike fields} lead to serious problems with
respect to the activation of (compactly localized) particle counters.

In the old days \cite{Weinbook} infinite spin representations were rejected on
the ground that nature does not make use of them. But whether nowadays, i.e.
in times of dark matter, one would uphold such dismissals is questionable.
String-localized quantum fields fluctuate both in $x$ as well as in
$e$\footnote{These long distance (infrared) fluctuations are short distance
fluctuation in the sense of the asymptotically associated d=1+2 de Sitter
spacetime.}. The can always be constructed in such a way that their effective
short distance dimension is the lowest possible one allowed by positivity,
namely $d_{sd}=1$ for all spins$.$It is very difficult to construct the
covariant "infinite spin" fields by the group theoretic intertwiner method
used by Weinberg \cite{Weinbook}; in \cite{MSY} the more powerful setting of
modular localization was used.

For pointlike generating fields $\Psi^{(A,\dot{B})}(x)$ one finds the
following two relations between the physical spin (helicity) and the possible
range of spinorial indices%

\begin{align}
&  \left\vert A-\dot{B}\right\vert \leqslant s\leqslant A+\dot{B}%
\label{spin}\\
&  h=A-\dot{B},~m=0\nonumber
\end{align}
In the massive case all possibilities for the angular decomposition of two
spinorial indices are allowed whereas in the massless case the values of the
helicity $h$ are severely restricted (second line). For $(m=0,h=1)$ the
formula reproduces the impossibility of reconciling pointlike vector
potentials with the Hilbert space positivity. This holds for all
$(m=0,s\geq0):$ pointlike localized "field strengths" (in h=2, the linearized
Riemann tensor) have no pointlike quantum "potentials" (in h=2, the $g_{\mu
\nu}$) and represents one solution of the famous clash between localization
and the Hilbert space structure. Since the classical theory does not care
about positivity, the (Lagrangian) quantization setting inevitably forces the
scarification of the Hilbert space in favor of Krein spaces (implemented by
the Gupta-Bleuler or BRST formalism). The more intrinsic Wigner
representation-theoretical approach keeps the Hilbert space and lifts the
unmotivated restriction to pointlike generators in favor of semiinfinite
stringlike generating fields.

For $(m=0,s=1)$ the stringlike covariant potentials $A_{\mu}(x,e)$ are
uniquely determined in terms of the field strength $F_{\mu\nu}(x)$ and a
spacelike direction $e.$ The idea is somewhat related to Mandelstam's attempt
to formulate QED without the vectorpotentials \cite{Mandel}. But even though
the string-local potential is uniquely determined in terms of $F_{\mu\nu},e,$
it is much safer to explicitly introduce the $A_{\mu}(x,e)$ because they are a
strong reminders that one is dealing with objects which fluctuate in both $x$
and $e;$ in fact the improvement of the short distance property in $x$ is paid
for by a worsening infrared behavior i.e. the $A_{\mu}(x,e)$ is an
operator-valued distribution in both $x,e$. In contrast to the above infinite
spin representation which cannot be pointlike generated, all other
representations admit pointlike generators and only exclude pointlike potentials.

As an illustrative example let us look at the Aharonov-Bohm effect in
QFT\footnote{The standard A-B effect is about quantum mechanical charged
particle in an \textit{external} magnetic field.}. In terms of Haag's
intrinsic LQP setting of QFT this is a breakdown of \textit{Haag duality} for
a toroidal spacetime localization \cite{charge}\cite{nonlocal}%
\begin{align}
\mathcal{A}(\mathcal{T}^{\prime\prime})  &  \varsubsetneq\mathcal{A}%
(\mathcal{T})^{\prime\prime}\\
\mathcal{T~}~spatial~torus~at~t  &  =0~,~\mathcal{T~}^{\prime\prime}\text{
}its~causal~completion\nonumber
\end{align}
For lower spin zero mass fields or for a torus-localized algebra from a
massive field of any spin one finds the equality sign (Haag duality). This can
be shown in terms of field strengths, but if one (for the convenience of
applying Stokes theorem) uses potentials it is easy to see that the indefinite
metric potential leads to the wrong result (zero effect) whereas the
string-localized potential in the Hilbert space accounts correctly for the A-B effect.

In massive theories there is no such clash; pointlike potentials of field
strength exist, but their short distance dimensions increase just like those
of field strengths. Nevertheless one can introduce stringlike potentials as a
means to lower the short distance dimension in order make couplings fit for
renormalization. The connection between the stringlike vectorpotential and its
pointlike counterpart (the $d_{sd}=2~$Proca field) leads to a scalar
string-localized field (all relations take place in Hilbert space)%
\begin{equation}
A_{\mu}(x,e)=A_{\mu}^{P}(x)+\partial_{\mu}\phi(x,e)
\end{equation}

The strategy for calculations of correlations in e.g. QED is then the
following. Use the $d_{sd}=1$ string fields for the perturbative calculations
in \textit{massive} QED. If needed, pass to the pointlike Proca field in every
order. But the pointlike Proca field has no zero mass limit (the Hilbert
space-localization clash), only the string-localized massive potential passes
to its zero mass counterpart.

For the charge-carrying matter fields the counterpart of the additive change
from string-localized to point-localized fields is multiplicative. The massive
Dirac-charge-carrying string-localized spinor of massive QED\footnote{The
perturbative transfer of string-localization from the vectorpotential leads to
a physical string-localized Dirac field which carries Dirac+Maxwell charge.}
is then expected to pass to the Dirac+Maxwell charge-carrying "infraparticle"
field, whereas the pointlike matter field corresponds to the pointlike Proca potential.

The Krein space-based BRST setting of gauge theory has a more limited range.
Physical charged fields and their (off-shell) correlations \textit{are not
part of the BRST formalism}. Electrically charged particles appear in a
somewhat indirect way in form of a prescription for photon-inclusive cross
section which, unlike the LSZ reduction formalism, has no direct relation to
spacetime correlation functions. There is a new setting which takes up an old
problem \cite{Haag} concerning the use of the restriction of algebras to the
forward light cone\footnote{For massive theories this restriction maintains
the full information whereas for QED it leads to a natural (geometric)
infrared cutoff.} which casts additional conceptual light on this problem
\cite{BDR}.

The more interesting case is that of selfinteracting massive vectormesons.
Here the systematic application of Scharf's version \cite{Scharf} of operator
gauge invariance within the BRST setting implements the group symmetry; as
already understood by Stora, the gauge group structure does not have to be
postulated, it is fixed by other consistency requirements. Different from
massive QED, the consistency of the massive BRST formalism also requires the
presence of a chargeless scalar field, but without any spontaneous symmetry
breaking\footnote{This confirms the correctness of Swieca's viewpoint of a
Schwinger-Higgs charge screening mechanism instead of a Goldstone symmetry
breaking in which the massless Goldstone boson is subsequently swallowd in a
process which converts the photon into a massive vectormeson.}.

The remaining doubts about whether the presence of a Higgs field is a feature
of the BRST quantization formalism (the intermediate use of a Krein setting is
imposed by Lagrangian quantization and does not belong to the intrinsic
properties of the desired QFT) or a consequence of foundational principles can
only be solved in the string-localized setting of this problem. There are
other important problems for which one needs this formulation. A derivation of
asymptotic freedom from a low order dimensionally regularized beta function is
only conceptually acceptable if the beta function is part of a parametric
Callen-Symanzik equation; a beta all by itself is a meaningless global
quantity. However the derivation of the latter requires the existence of a
massive perturbation theory; the prototype of such a computation is the
massive Thirring model \cite{Go-Lo} where $\beta=0$ to all orders (thus
preempting the conformal invariance for vanishing mass). In massless Y-M
models even the off-shell correlations are infrared divergent in all covariant
gauges. This situation, which is usually blamed on an imagined nonperturbative
long-distance behavior (related to confinement) may actually be the undesired
consequence of imposing point-localization in a situation which really
requires string-localization.

The crucial problem is the reformulation of the iterative Epstein-Glaser
renormalization in terms of string-localized fields. This is particular tricky
in massive Y-M interactions where, in contrast to massive QED, the interaction
involves several strings. A first incomplete account of these problems was
given in \cite{Rio}, but meanwhile this technical aspect of what replaces the
diagonal of the pointlike iteration in case of strings has been solved
\cite{Jens}. Therefore we hope to be able to present an account of low order
perturbative calculations in the near future \cite{future}.

As mentioned in the previous section, the need to understand another more
hidden side of local quantum physics did not arise with the appearance of the
dual model and string theory, but it already existed in Jordan's first (1926)
model of a QFT \cite{E-J}. The thermal aspects arising from restricting the
QFT vacuum to a spacetime subregion belong to that "other side of QFT" which
the standard formalism does not really reveal; for this reason the thermal
aspects of the Einstein-Jordan conundrum remained for a long time unexplained
or where mistakenly thought to be caused by curvature in gravity theory. They
were only understood in a different more algebraic formulation QFT referred to
as local quantum physics (LQP), which places the modular localization property
into the center stage. Since this setting will not only be useful for the
construction of the previously mentioned string-localized fields but even more
important to understand the conceptual origin of particle crossing in the next
section, we will use the remainder of this section to present some elementary
facts about it.

It has been realized, first in a special context in \cite{S2}, and then in a
general mathematical rigorous setting in \cite{BGL} (see also \cite{Fa-Sc}%
\cite{MSY}), that there exists a \textit{natural localization structure} on
the Wigner representation space for any positive energy representation of the
proper Poincar\'{e} group. A convenient presentation can be given in the
context of spinless chargeless particle for which the $(m>0,s=0)$ Wigner
one-particle space is the Hilbert space $H_{1}$ of (momentum space) wave
functions with the inner product
\begin{equation}
\left(  \varphi_{1},\varphi_{2}\right)  =\int\bar{\varphi}_{1}(p)\varphi
_{2}(p)\frac{d^{3}p}{2p_{0}},~~\hat{\varphi}(x)=\frac{1}{\left(  2\pi\right)
^{\frac{3}{2}}}\int e^{-ipx}\varphi(p)\frac{d^{3}p}{2p_{0}}%
\end{equation}
In this case the covariant x-space amplitude is simply the on-shell Fourier
transform of this wave function whereas for $(m\geq0;s\geq1/2)$ the covariant
spacetime wave function is more involved as a consequence of the presence of
intertwiners $u(p,s)$ between the manifestly unitary and the covariant form of
the representation \cite{Weinbook}.

Selecting a wedge region e.g. $W_{0}=\{x\in\mathbb{R}^{d},x^{d-1}>\left\vert
x^{0}\right\vert \},$ one notices that the unitary wedge-preserving boost
$U(\Lambda_{W}(\chi=-2\pi t))=\Delta^{it}$ commutes with the antiunitary
reflection $J_{W}$ on the edge of the wedge (i.e. along the coordinates
$x^{d-1}-x^{0}$). The distinguished role of the wedge region is that they form
a \textit{commuting pair} of (boost, antiunitary reflection). This has the
unusual (and perhaps even unexpected) consequence that the unbounded and
antilinear operator%
\begin{align}
S_{W}  &  :=J_{W}\Delta^{\frac{1}{2}},~~S_{W}^{2}\subset1\\
&  since~~J\Delta^{\frac{1}{2}}J=\Delta^{-\frac{1}{2}}\nonumber
\end{align}
which is intrinsically defined in terms of Wigner representation data, is
\textit{involutive on its dense domain} and therefore has a unique closure
with $ranS=domS~$(unchanged notation for the closure).

The involutivity means that the $S$-operator has $\pm1$ eigenspaces; since it
is antilinear, the +space multiplied with $i$ changes the sign and becomes the
- space; hence it suffices to introduce a notation for just one eigenspace%
\begin{align}
K(W)  &  =\{domain~of~\Delta_{W}^{\frac{1}{2}},~S_{W}\psi=\psi\}\\
J_{W}K(W)  &  =K(W^{\prime})=K(W)^{\prime},\text{ }duality\nonumber\\
\overline{K(W)+iK(W)}  &  =H_{1},\text{ }K(W)\cap iK(W)=0\nonumber
\end{align}

It is important to be aware that we are dealing here with \textit{real}
(closed) subspaces $K$ of the complex one-particle Wigner representation space
$H_{1}$. An alternative is to directly work with the complex dense subspaces
$K(W)+iK(W)$ as in the third line. Introducing the \textit{graph norm} in
terms of the positive operator$~\Delta,$ the dense complex subspace becomes a
Hilbert space $H_{1,\Delta}$ in its own right. The upper dash on regions
denotes the causal disjoint (the opposite wedge), whereas the dash on real
subspaces means the symplectic complement with respect to the symplectic form
$Im(\cdot,\cdot)$ on $H.$ All the definition work for arbitrary positive
energy representations of the Poincare group.

The two properties in the third line are the defining relations of what is
called the \textit{standardness property} of a real
subspace\footnote{According to the Reeh-Schlieder theorem a local algebra
$\mathcal{A(O})$ in QFT is in standard position with respect to the vacuum
i.e. it acts on the vacuum in a cyclic and separating manner. The spatial
standardness, which follows directly from Wigner representation theory, is
just the one-particle projection of the Reeh-Schlieder property.}; any
abstract standard subspace K of an arbitrary real Hilbert with a $K$-operator
space permits to define an abstract $S$-operator in its complexified Hilbert
space%
\begin{align}
&  S(\psi+i\varphi)=\psi-i\varphi,~S=J\Delta^{\frac{1}{2}}\label{inv}\\
&  domS=dom\Delta^{\frac{1}{2}}=K+iK\nonumber
\end{align}
whose polar decomposition (written in the second line) yields two modular
objects, a unitary modular group $\Delta^{it}$ and an antiunitary reflection
which generally have however no geometric interpretation in terms of
localization. The domain of the Tomita $S$-operator is the same as the domain
of $\Delta^{\frac{1}{2}},$ namely the real sum of the $K$ space and its
imaginary multiple. Note that for the physical case at hand, this domain is
intrinsically determined solely in terms of the Wigner group representation theory.

The $K$-spaces are the real parts of these complex $domS,$ and in contrast to
the complex domain spaces they are closed as real subspaces of the Hilbert
space (corresponding to the one-particle projection of the real subspaces
generated by Hermitian Segal field operators).$~$Their symplectic complement
can be written in terms of the action of the $J~$operator and leads to the
K-space of the causal disjoint wedge $W^{\prime}$ (Haag duality)%
\begin{equation}
K_{W}^{\prime}:=\{\chi|~Im(\chi,\varphi)=0,all\text{ }\varphi\in K_{W}%
\}=J_{W}K_{W}=K_{W^{\prime}}%
\end{equation}

The extension of W-localization to arbitrary spacetime regions $\mathcal{O}$
is done by representing the causal closure $\mathcal{O}^{\prime\prime}$ as an
intersection of wedges and defining $K_{\mathcal{O}}$ as the corresponding
intersection of wedge spaces
\begin{equation}
K_{\mathcal{O}}=K_{\mathcal{O}^{\prime\prime}}\equiv%
{\displaystyle\bigcap\limits_{W\supset\mathcal{O}^{\prime\prime}}}
K_{W},~~\mathcal{O}^{\prime\prime}=causal~completion~of~\mathcal{O} \label{K}%
\end{equation}
These K-spaces lead via (\ref{inv}) and (\ref{K}) to the modular operators
associated with $K_{\mathcal{O}}.$

For those who are familiar with Weinberg's intertwiner formalism
\cite{Weinbook} for passing from the unitary Wigner to covariant
representations in the dotted/undotted spinor formalism, it may be helpful to
recall the resulting "master formula"%

\begin{align}
&  \Psi^{(A,\dot{B})}(x)=\frac{1}{\left(  2\pi\right)  ^{\frac{3}{2}}}%
\int(e^{-ipx}\sum_{s_{3}=\pm s}u^{(A,\dot{B})}(p,s_{3})a(p,s_{3}%
)+\label{field}\\
&  \ \ \ \ \ \ \ \ \ \ \ \ \ \ \ \ +e^{ipx}\sum_{s_{3}=\pm s}v^{(A,\dot{B}%
)}(p,s_{3})b^{\ast}(p,s_{3}))\frac{d^{3}p}{2\omega}\nonumber\\
&  \sum_{s_{3}=\pm s}u^{(A,\dot{B})}(p,s_{3})a(p,s_{3})\rightarrow u(p,e)\cdot
a(p)
\end{align}
where the a,b amplitudes correspond to the Wigner momentum space wave
functions of particles/antiparticles and the u,v represent the intertwiner and
its charge conjugate. For the third class (infinite spin, last line) the sum
over spin components has to be replaced by an inner product between a $p,e$
dependent infinite component intertwiner $u$ and an infinite component $a(p),$
because in this case Wigner's "little space" is infinite dimensional. The
$\Psi(x)$ respectively $\Psi(x,e)$ are "generating wave functions" i.e. they
are wavefunction-valued Schwartz distributions which by smearing with
$\mathcal{O}$-supported test functions become $\mathcal{O}$-localized wave
functions. Adding the opposite frequency anti-particle part one obtains the
above formula which by re-interpreting the $a^{\#},b^{\#}$ as
creation/annihilation operators (second quantization) become
point-respectively string- like free fields. The second quantization functor
maps the complex amplitudes $a,b$ into creation/annihilation operators. The
resulting operator-valued Schwartz distributions are global objects
(generators) in the sense that they generate $\mathcal{O}$-localized operators
$\Psi(f)$ by "smearing" them with $\mathcal{O}$-supported test functions
$suppf\in\mathcal{O}.$

Only the massive case the full spectrum of spinorial indices $A,\dot{B}$ is
exhausted (\ref{spin}) whereas the massless case leads to huge gaps which come
about because pointlike "field-strength" are allowed whereas pointlike
"potentials" are rejected. With the awareness about the conceptual clash
between localization and the Hilbert space\footnote{In the case of \cite{MSY}
this awareness came from the prior use of "modular localization" starting in
\cite{S1}\cite{S2} but foremost (covering $\mathit{all}$ positive energy
Wigner representations) in \cite{BGL}.}.

The difference to Weinberg's setting is that, whereas he uses the
computational somewhat easier manageable covariance requirement (for wave
functions and free fields covariance is synonymous with causal localization,
but in the presence of interaction the localization of operators and that of
states split apart), the modular localization method uses causal localization
directly and bypasses the issue of the nonunique intertwiners by aiming
directly at "modular-localized" dense subspaces.

The generating pointlike fields are indispensable in the implementation of
perturbation theory. They are the mediators between classical localization
(which is used when one specifies zero order interactions in form of invariant
Wick-ordered polynomials) and quantum localization, which takes over when one
uses the Epstein-Glaser iteration machinery to implement the causal
localization principle order by order \cite{E-G} . Modular localization on the
other hand is essential in trying to \textit{generalize Wigner's intrinsic
representation theoretical approach to the} (non-perturbative) \textit{realm
of interacting localized observable algebras} (next section).

In order to arrive at Haag's setting of local quantum physics in the absence
of interactions, one only has to apply the Weyl functor $\Gamma$ which maps
wave functions into operators and wave function spaces into operator algebras
(or its fermionic counterpart), symbolically indicated by the functorial
relation%
\begin{equation}
K_{\mathcal{O}}\overset{\Gamma}{\rightarrow}\mathcal{A(O}) \label{func}%
\end{equation}
The functorial map $\Gamma~$\textit{also}$~$relates\ the modular operators
$S,J,\Delta$ from the Wigner wave function setting directly with their "second
quantized" counterparts $S_{Fock},J_{Fock},\Delta_{Fock}$ in Wigner-Fock
space; it is then straightforward to check that they are precisely the modular
operators of the Tomita-Takesaki modular theory applied to causally localized
operator algebras.%

\begin{align}
\sigma_{t}(\mathcal{A(O}))  &  \equiv\Delta^{it}\mathcal{A(O})\Delta
^{-it}=\mathcal{A(O})\label{sig}\\
J\mathcal{A(O})J  &  =\mathcal{A(O})^{\prime}=\mathcal{A(O}^{\prime})\nonumber
\end{align}
In the absence of interactions these operator relation are consequences of the
modular relations for Wigner representations. The Tomita-Takesaki theory
secures their general existence for \textit{standard pairs} ($A,\Omega$) i.e.
an operator algebras $\mathcal{A}$ and a state vector $\Omega\in H$ on which
$\mathcal{A}$ acts cyclic and separating (no annihilators of $\Omega$ in
$\mathcal{A}$). The polar decomposition of the antilinear closed Tomita
$S$-operator leads to the unitary modular automorphism group $\Delta^{it}%
~$associated with~the subalgebra $\mathcal{A(O})\subset B(H)$ and the vacuum
state vector $\Omega$ i.e. with the pair $(\mathcal{A(O}),\Omega).$

Although $B(H)$ is generated from the two commuting algebras $\mathcal{A(O})$
and $\mathcal{A(O}^{\prime}),$ they do not form a tensor product in $B(H)$;
hence the standard quantum-information concepts concerning entaglement and
density matrices are not applicable. In contrast to QM where one has to
average over degrees of freedom in order to convert entangled states into
density matrices, modular situations are distinguished in that the averaging
is replaced by the trivial operation of just restricting the global "standard"
state (e.g. the vacuum) to the local subalgebra of interest. $~$

The only case for which the modular localization theory (the adaptation of the
Tomita-Takesaki modular theory to the causal localization principle of QFT)
has a geometric interpretation, independent of whether interactions are
present or not and independent of the type of quantum matter, is the wedge
region i.e. the Lorentz transforms of the standard wedge $W=\left\{
x_{0}<x_{3}|\mathbf{x}_{tr}\in\mathbb{R}^{2}\right\}  .~$In that case the
modular group is the wedge-preserving Lorentz boost and the $J$ represents a
reflection on the edge of the wedge i.e. it is up to a $\pi$-rotation equal to
the antiunitary TCP operator. The derivation of the TCP invariance as derived
by Jost \cite{Jost}, together with scattering theory (the TCP transformation
of the S-matrix) leads to the relation
\begin{equation}
J=S_{scat}J_{in} \label{jost}%
\end{equation}
which in \cite{S1}\cite{S2} has been applied to constructive problems of
integrable QFTs. The is a relation which goes much beyond scattering theory;
in fact it only holds in local quantum physics since it attributes the new
role of a relative modular invariant of causal localization to the S-matrix.

This opens an unexpected fantastic new possibility of a \textit{top to bottom
construction} of QFT in which the first step is the construction of generators
for the wedge-localized algebra $\mathcal{A}(W)$ and the sharpening of
localization is done by intersecting wedge algebras. Compact localized double
cone algebras and their generating pointlike fields would only appear at the
end. In fact according to the underlying philosophy that all relevant physical
data can be obtained from localized algebras, thus avoiding the use of
individual operators within such an algebra. This is the tenor of the paper
"\textit{On revolutionizing quantum field theory with Tomita's modular
theory"~}\cite{Bo} by Borchers, to whose memory I have dedicated this
paper\footnote{Please note that the word "revolution" in this context has a
completely different meaning from its use in string theory.}. The next section
presents the first step in such a construction.

The only prerequisites for the general (abstract) case is the "standardness"
of the pair ($\mathcal{A},\Omega$) where "standard" in the theory of operator
algebras means that $\Omega$ is a cyclic and separating vector with respect to
$\mathcal{A}$, a property \textit{which in QFT is always fulfilled} for
localized $\mathcal{A(O})^{\prime}s,$ thanks to the validity of the
Reeh-Schlieder theorem \cite{Haag}. These local operator algebras of QFT are
what I referred to in previous publications as a \textit{monad;} there
properties are remarkably different from the algebra of all bounded operators
$B(H)$ which one encounters for Born-localized algebras \cite{interface}. For
general localization regions the modular unitaries have no geometric
interpretation (they describe a kind of fuzzy action inside $\mathcal{O}$) but
they are uniquely determined in terms of intersections of their geometric
$W$-counterparts, a top to bottom strategy which is quite efficient, even in
the simpler context of localized subspaces $K_{\mathcal{O}}~\ $related to
Wigner's positive energy representation theory for the Poincar\'{e} group
\cite{BGL}.

The most important conceptual contribution of modular localization theory in
the context of the present work is the assertion that the reduction of the
global vacuum (and also finite energy particle states) to a local operator
algebra $\mathcal{A(O})$ leads to a thermal state for which the "thermal
Hamiltonian" $H_{mod}$ is the generator of the modular unitary group%
\begin{align}
&  e^{-i\tau H_{mod}}:=\Delta^{i\tau}\label{mod}\\
&  \left\langle AB\right\rangle =\left\langle Be^{-H_{mod}}A\right\rangle
\nonumber
\end{align}
where the second line is what one obtains for heat bath thermal systems after
rewriting the Gibbs trace formula into the state-setting of the \textit{open
system formulation} of statistical mechanics \cite{Haag}. Whereas the trace
formulation breaks down in the thermodynamic limit, this analytic KMS formula
(asserting analyticity in $-1<Im\tau<0$) remains. It is in this and only in
this limit, that QM produces a global monad algebra (different from $B(H))$
which is of the same type as the localized monad of QFT.

This underlines again the intimate connection between quantum causal
localization and the ensemble nature of measurements in QFT (further remarks
in the last section). Note that it is of cause not forbidden to speak about
concrete operators in $\mathcal{A(O})$, the main difference of the ensemble
viewpoint of QFT and the attribution of probabilities to individual events (as
a result of the absence of an intrinsic mechanism which requires ensembles) is
that the former tries to extract the description of nature from properties of
existing localized ensembles. For the case at hand this is the modular
Hamiltonian $H_{mod}$ which changes together with the standard pair
($\mathcal{A(O}),\Omega$) in such a way that the Hamiltonian for the algebra
with the larger localization can also be applied to an algebra localized
inside a causally closed region, but not the other way around. This leads to
an infinite supply of modular Hamiltonians which all live in the same Hilbert
space; this incredible rich structure has no counterpart in QM.

As well-known Einstein had serious problems with the assignment of
probabilities to single events as usually (not by everybody) done in Born's
probabilistic interpretation of QM. Since QFT is more foundational than QM one
should perhaps consider the extrinsic probability of global QM as a limiting
case of the ensemble interpretation in which the thermal aspects of modular
localization and vacuum polarization get lost. It can be assumed that Einstein
would have accepted the thermal probability arising from localization in QFT
if it would have been available during his time .

Closely related is the "GPS" characterization of a QFT, including its
Poincar\'{e} spacetime symmetry as well as the internal symmetries of its
quantum matter content, in terms of modular positioning of a finite number of
monads in a shared Hilbert space. For d=1+1 chiral models the number of monads
is 2 or 3, depending on the formulation whereas in d=1+3 the smallest number
for a GPS construction is 7. This way of looking at QFT is an extreme
relational point of view in terms of objects which have no internal structure;
this explains the terminology "monad" (a realization of Leibnitz point of view
in the context of abstract quantum matter) \cite{W-K}\cite{interface}. As life
is an holistic phenomenon since it cannot be explained from its chemical
ingredients so is QFT which cannot be understood in terms of properties of a
monad. This is a philosophical view of QFT which exposes its radically
holistic structure in the most forceful way; in praxis one starts with one
monad and assumes that one knows the action of the Poincar\'{e} group on it
\cite{S1}\cite{S2}; this was the way in which the existence of factorizing
models was shown \cite{Lech}.

As mentioned in the introduction and more forceful in the last section, the
intrinsic thermal aspect of localization is the reason why the probability
issue in QFT is conceptually radically different from QM for which Born
localization does not lead to a probability; the latter rather has to be added.

Although the functorial relation between the Wigner theory and operator
algebras breaks down in the presence of (any) interactions, there is a weak
substitute called "emulation" (it emulates W-smeared free field $\Psi(f)$
inside the interacting $\mathcal{A}(W)$). It is extremely powerful in terms of
integrable systems and promises to have clout even outside this special
family; this will be the main topic of next section.

The modular analysis has some simple consequences about the issue of string
localization. There is a whole family of Wigner representations (the infinite
spin family) for which the intersections $K_{\mathcal{O}}$ vanish for compact
$\mathcal{O}$ but not for $\mathcal{O}=\mathcal{C}~$a$~$spacelike cone
\cite{BBS}. This is the origin of the spacelike string generators for
spacelike cone localized subspaces \cite{MSY}. The upshot is the existence of
generating fields $\Psi^{(A,\dot{B})}(x;e)$ which are localized on the
semiinfinite line $x+\mathbb{R}_{+}e$ and fluctuate both in $x$ \textit{and}
$e.~$Their perturbative use requires a nontrivial extension of the
Epstein-Glaser approach. An important "fringe benefit" of the use of
string-localized potentials is that the best (smallest allowed by positivity)
short distance dimension namely $d_{sd}=1~$can always be attained by the use
of suitable potentials instead of field strengths (whose $d_{sc}$ increases
with $s$).

This property is preserved in the massive case, although in this case there is
no representation-theoretic reason for using such $A_{\mu}(x,e)$ potentials.
The standard pointlike massive potentials $A_{\mu}(x)~$have $d_{sc}=2$; it is
only the stringlike potentials which allow a smooth transition in the limit
$m\rightarrow0.$ Whereas for interactions in terms of pointlike fields there
is exists in d=1+3 only a finite number of interactions which stay within the
power-counting limit, this limit allows an infinite set of couplings with the
help of string-localized fields. Using this new setting, many of the
unanswered problems of the gauge theoretic setting (the Higgs issue) hopefully
will be laid to rest \cite{future}.

The problems of gauge theory are very much related to the mentioned clash
between pointlike generators with the Hilbert space structure of QT. Although
in this paper we focus mainly on thermal manifestations of localization and
crossing properties, issues of gauge theories and asymptotic freedom also
depend strongly on causal localization, more than most particle physicists
might have hitherto imagined. The history of that issue did not start with the
famous Politzer-Gross-Wilczek work, but had its precursor in the observation
by Parisi and Symanzik that sign of the beta function changes if one inverts
the sign in the $A^{4}$ self-coupling.

In that case the computation based on the Callen Symanzik equation is
straightforward and since the theory is massive there is no infrared problem
which presents a perturbative derivation of the C-S equation. The latter is a
parametric differential equation for spacetime correlation functions of
pointlike fields whose physical content expresses how a change of coupling
constants can be compensated by the change of the other parameters in such a
way that one stays inside the finite parametric "island". It leads to a group
which only change the parametrization, but not the island itself, and the
Callen-Symanzik equation is the differential form of the renormalization
group. The Parisi-Symanzik demonstration of asymptotic freedom is very clear,
but unfortunately it is a toy model without any physical content.

For scalar fields as they appear in the formulation of critical phenomena it
computational quite efficient to follow Wilson and use the so-called
dimensional regularization method. This method is based on the formal idea
that scalar fields look the same in different spacetime dimensions (the
representation of the Wigner "little group" which determines the
spinorial/tensorial character is trivial) so that the existence of a smooth
interpolation has a certain plausibility. However for $s\geq1$ fields depend
in an essential way on the spacetime dimension.

When one applies this method to Yang-Mills gluons there are two obstacles:

\begin{itemize}
\item Renormalized correlation functions in the pointlike gauge theoretic
setting of Y-M theories have infrared singularities whose physical origin is
blamed on the not understood issue of confinement; in contrast to QED these
divergencies occur even off mass-shell in all covariant gauges. In a
string-localized physical (Hilbert space) setting these divergencies result
from fluctuations in the string-direction $e$ and therefore can be controlled
by smearing over spacelike string directions (points in d=1+2 de Sitter
space). It is precisely these fluctuations which reduce the short distance
dimension of vectorpotentials from 2 to 1. (in fact in that description the
infrared divergencies are short distance divergencies in the d=1+2 de Sitter
space of spacelike string directions). Since gauge invariant observables are
identical to pointlike localized subobservables within the stringlike setting,
a crucial test whether infrared divergencies have their origin in the gauge
theoretic treatment of gluons or are fundamentally nonperturbative would be
the calculation of pointlike composites $\sum_{a}\mathcal{N(}F_{\mu\nu}%
^{a}(x)F_{\kappa\lambda}^{a}(x))~$where $\mathcal{N}$ denotes the "normal
product". In QED the on-shell infrared divergencies do have their origin in
the transfer of semiinfinite string-localization from the stringlike
potentials to the charged spinor matter whereas in Y-M theories there is no
distinction between the string-localized transmitters (gluons) of interactions
and the objects which suffer the interaction (also gluons). One expects a much
stronger stringlike localization from a perturbation theory based on the zero
mass limiting behavior of massive stringlike Yang-Mills models.

\item Even if the infrared problem is solved and renormalized perturbative
correlations of gauge-invariant composites fulfilling C-S parametric
differential equations exist, the use of dimensional regularization still
remains somewhat questionable from the viewpoint of localization for $s\geq1$
since one is not in Wilson's situation of critical phenomena which are
described in terms of scalar fields to which the intuitive idea of a
smoothness in the spacetime dimensionality does apply. Already the Wigner
representation theory for $s\geq1~$depends (through the "little group)
significantly on spacetime. To the extend that the renormalization can be done
by other regularization methods or without regularization a l\'{a}
Epstein-Glaser, this caveat is irrelevant.
\end{itemize}

The asymptotic freedom calculation (where only the sign in beta is important)
by Politzer and Gross-Wilczek has led to successful experimental verifications
and is the basis of theoretical precision calculations. From a conceptual
point of view the situation has maintained a circular albeit selfconsistent
aspect: the beta function by itself (i.e. without correlation functions and
the C-S equations of which it is a part) is incomplete and the precision
calculations for correlations (which use the perturbative beta function in
order to describe the short distance behavior) use the opposite sign from this
incomplete result. Both, the infrared behavior and the beta function problem
are tied to a deeper understanding of localization for $s\geq1$.

It was the problem of localization for Wigner's infinite spin class of
positive energy representations of the Poincar\'{e} group which directed the
attention to the issue of string localization \cite{BGL}\cite{MSY}. The reason
why this important issue was discovered rather late is that it does not permit
a Lagrangian characterization; even Weinberg's method of covariant
intertwiners based on Wigner's representation theory encountered difficulties
in "covariantizing" these representations; the appropriate method for their
construction (first of their $K_{\mathcal{C}}$ -spaces and then their
string-localized generating wave functions) is modular localization. On the
other hand the predominant method in QFT has been Lagrangian/functional
quantization which has no access to string-localization.

A string field in the sense of ST has no relation with string-localized
fields. "String" in ST refers to the classical Polyakov Lagrangian which
contains \textit{classical stringlike objects} $X_{\mu}(\sigma,\tau).$ From
the impossibilty to understand relativistic quantum particles by quantizing
$L\symbol{126}\sqrt{ds^{2}},$ one should be deeply suspicious about
attributing the word "string" to an object on the basis of a questionable
quantization; as a covariant particle description cannot be obtained by
quantizing this classical relativistic particle Lagrangian it would be foolish
to expect a quantum string to arise from the quantization of the Nambu-Goto action.

As genuinely string-localized objects cannot be obtained by quantization (but
rather by using Wigner's representation theory), objects obtained by
quantization (as ST) are not string-localized in the quantum sense. The
unmanageable infrared divergencies in the gauge-theoretic setting of
Yang-Mills theory are a reminder of the presence of the noncompact
string-localization of vectormesons. The reference of string theorists to
strings in the sense of gauge bridges between opposite charges is misleading,
there are no quantum strings since ST is pointlike generated; this holds not
only for the standard ST, but also for recent attempts to free the Nambu-Goto
action from the nonsensical spacetime dimensional restriction \cite{BRZ}.

\section{Generators of wedge algebras, "Wignerism" in the presence of
interactions}

The basic idea which underlies the new setting of QFT is to avoid quantization
and follow instead Wigner's representation theoretical setting. As explained
in the previous section, this approach leads to free fields in two steps: the
classification of positive energy representations of the Poincar\'{e} group,
and its use in a functorial setting (second quantization functor) which maps
modular localized real subspaces into localized operator algebras (or
pointlike wave functions into quantum fields). The main computational work is
the classification; knowing modular localization the second functorial step is
self-directed. In this way particle state vectors and state vectors obtained
by applying free fields to the vacuum become synonymous.

It is well-known that this direct particle-field relation \footnote{The
particle-field relation through scattering theory is asymptotic ; here we are
interested in relations within localized regions of spacetime.} breaks down in
the presence of \textit{any interaction}. The following theorem shows that the
separation between the two is very drastic indeed:

\begin{theorem}
(Mund's algebraic extension \cite{Mund2} of the old J-S theorem \cite{St-Wi})
A Poincar\'{e}-covariant QFT in $d\geq1+3$ fulfilling the mass-gap hypothesis
and containing a \textit{sufficiently large} set of "temperate" wedge-like
localized vacuum polarization-free one-particle generators (PFGs) is unitarily
equivalent to a free field theory.
\end{theorem}

The only relic of the functorial relation which remains unaffected by this
theorem is a rather weak relation between particles and local fields in
wedge-localized regions. The idea is to obtain a kind of \textit{"emulation"
of free incoming fields} (\symbol{126}particles) restricted to a wedge regions
\textit{inside the interacting wedge algebra} as a replacement for the
nonexisting second quantization functor. This is achieved with the help of
modular localization theory.

The starting point is a \textit{bijection} between wedge-localized incoming
fields operators and interacting operators. This bijection is based on the
equality of the dense subspace which these operators from the two different
algebras create from the vacuum. Since the domain of the Tomita $S$ operators
for two algebras which share the same modular unitary $\Delta^{it}$ is the
same, a vector $\eta\in domS\equiv domS_{\mathcal{A}(W)}=dom\Delta^{\frac
{1}{2}}~$is also in $domS_{\mathcal{A}_{in}(W)}=\Delta^{\frac{1}{2}}$ (in
\cite{BBS} it was used for one-particle states). In more explicit notation,
which emphasizes the bijective nature, one has
\begin{align}
A\left\vert 0\right\rangle  &  =A_{\mathcal{A(}W\mathcal{)}}\left\vert
0\right\rangle ,~A\in\mathcal{A}_{in}(W),~A_{\mathcal{A(}W\mathcal{)}}%
\in\mathcal{A}(W)\\
S(A)_{\mathcal{A(}W\mathcal{)}}\left\vert 0\right\rangle  &  =(A_{\mathcal{A(}%
W\mathcal{)}})^{\ast}\left\vert 0\right\rangle =S_{scat}A^{\ast}S_{scat}%
^{-1}\left\vert 0\right\rangle ,~S=S_{scat}S_{in}~\nonumber\\
&  S_{scat}A^{\ast}S_{scat}^{-1}\in\mathcal{A}_{out}(W)
\end{align}
Here $A$ is either an operator from the wedge localized free field operator
algebra $\mathcal{A}_{in}(W)$ or an (unbounded) operator affiliated with this
algebra (e.g. products of incoming free fields $A(f)$ smeared with
$f,~suppf\in W$); $S$ denotes the Tomita operator of the interacting algebra
$\mathcal{A}(W)$. Under the assumption that the dense set generated by the
dual wedge algebra $\mathcal{A}(W)^{\prime}\left\vert 0\right\rangle $ is in
the domain of definition of the bijective defined "emulats" (of the
wedge-localized free field operators inside its interacting counterpart) the
$A_{\mathcal{A}(W)}$ are uniquely defined; in order to be able to use them for
the reconstruction of $\mathcal{A}(W)$ the domain should be a core for the
emulats. Unlike smeared Wightman fields, the emulats $A_{\mathcal{A(}%
W\mathcal{)}}$ do not define a polynomial algebra, since their unique
existence does not allow to impose additional properties; in fact they only
form a vector space and the associated algebras have to be constructed by
spectral theory or other means to extract an algebra from a vector space of
closed operators (as Connes reconstruction of an operator algebra from its
positive cone state structure).

Having settled the problem of uniqueness, the remaining task is to determine
their action on wedge-localized multi-particle vectors and to obtain explicit
formulas for their particle formfactors. All these problems have been solved
in case the domains of emulats are invariance under translations; in that case
the emulats possess a Fourier transform \cite{BBS}. This requirement is
extremely restrictive and is only compatible with d=1+1 elastic two-particle
scattering matrices of integrable models\footnote{This statement, which I owe
to Michael Karowski, is slightly stronger than that in \cite{BBS} in that that
higher elastic amplitudes are combinatorial products of two-particle
scattering functions, i.e. the only solutions are the factorizing models.}; in
fact it should be considered as the foundational definition of integrability
of QFT in terms of properties of wedge-localized generator \cite{integrable}.

Since the action of emulats on particle states is quite complicated, we will
return to this problem after explaining some more notation, formulating the
crossing identity in connection with its KMS counterpart and remind the reader
of how these properties have been derived in the integrable case.

For integrable models the wedge duality requirement (\ref{dual}) leads to a
unique solution (the Zamolodchikov-Faddeev algebra), whereas for the general
non-integrable case we will present arguments, which together with the
comparison with integrable case determine the action of emulates on particle
states. The main additional assumption is that the only way in which the
interaction enters the this construction of bijections is through the
S-matrix\footnote{A very reasonable assumption indeed because this is the only
interaction-dependent object which enters as a relative modular invariant the
modular theory for wedge localization.}. With this assumption the form of the
action of the operators $A_{\mathcal{A(}W\mathcal{)}}$ on multiparticle states
is fixed. The ultimate check of its correctness through the verification of
wedge duality (\ref{dual}) is left to future investigations.

Whereas domains of emulats in the integrable case are translation invariant
\cite{BBS}, the only domain property which is always preserved in the general
case is the invariance of the domain under the subgroup of those Poincar\'{e}
transformations which leave W invariant. In contrast to QM, for which
integrability occurs in any dimension, integrability in QFT is restricted to
d=1+1 factorizing models \cite{integrable}.

A basic fact in the derivation of the crossing identity, including its
analytic properties which are necessary in order to return to the physical
boundary, is the \textit{cyclic KMS property.} For three operators affiliated
with the interacting algebra $\mathcal{A}(W),$ two of them being emulates of
incoming operators\footnote{There exists also a "free" KMS identity in which
$B$ is replaced by $\left(  B\right)  _{\mathcal{A}_{in}(W)}$ so everything
refers to the algebra $\mathcal{A}_{in}(W).~$The derivation of the
corresponding crossing identity is rather simple and its use is limited to
problems of writing iterating fields as a series of Wick-ordered product of
free fields.} it reads:%

\begin{align}
&  \left\langle 0|BA_{\mathcal{A}(W)}^{(1)}A_{\mathcal{A}(W)}^{(2)}%
|0\right\rangle \overset{KMS(\mathcal{A}(W))}{=}\left\langle 0|A_{\mathcal{A}%
(W)}^{(2)}\Delta BA_{\mathcal{A}(W)}^{(1)}|0\right\rangle \label{M}\\
&  A^{(1)}\equiv:A(f_{1})...A(f_{k}):,~A_{in}^{(2)}\equiv:A(f_{k+1}%
)...A(f_{n}):,~suppf_{i}\in W\nonumber
\end{align}
where in the second line the operators were specialized to Wick-ordered
products of smeared free fields $A(f)$ which are then emulated within
$\mathcal{A}(W).~$Their use is necessary in order to convert the KMS relation
for $\mathcal{A}(W)$ into an identity of \textit{particle formfactors} of the
operator $B\in\mathcal{A}(W)$.\ If the bijective image acts on the vacuum, the
subscript $\mathcal{A}(W)~$for the emulats can be omitted and the resulting
Wick-ordered product of free fields acting on the vacuum describe a
multi-particle state in $\hat{f}_{i}$ momentum space wave functions. The roof
on top of $f$ denotes the wave function which results from the forward mass
shell restriction of the Fourier transform of W-supported test function. The
result are wave functions in a Hilbert space of the graph norm $\left(
\hat{f},\left(  \Delta+1\right)  \hat{f}\right)  $ which forces them to be
analytic in the strip $0<Im\theta<\pi.$

The derivation of the crossing relation requires to compute the formfactor of
the emulate $A_{\mathcal{A}(W)}^{(1)}$ between W-localized particle states and
a general W-localized state. For simplicity of notation we specialize to d=1+1
in which case neither the wedge nor the mass-shell momenta have a transverse
component and particles are characterized by their rapidity. Using the
analytic properties of the wave functions which connect the complex conjugate
of the antiparticle wave function with the $i\pi~$boundary value of the
particle wave function, one obtains%
\begin{align}
&  \int..\int\hat{f}_{1}(\theta_{1})...\hat{f}_{1}(\theta_{n})F^{(k)}%
(\theta_{1},...,\theta_{n})d\theta_{1}...d\theta_{n}=0\\
&  F^{(k)}(\theta_{1},..,\theta_{n})=\left\langle 0\left\vert BA_{\mathcal{A}%
(W)}^{(1)}(\theta_{1},..,\theta_{k})\right\vert \theta_{k+1},..,\theta
_{n}\right\rangle _{in}-\nonumber\\
~~~~~~~~~~~~~~~~~~~~  &  -~_{out}\left\langle \bar{\theta}_{k+1}%
,..,\bar{\theta}_{n}\left\vert \Delta^{\frac{1}{2}}B\right\vert \theta
_{1},..,\theta_{k}\right\rangle _{in}\nonumber
\end{align}
Here $\Delta^{\frac{1}{2}}$ of $\Delta~$was used to re-convert the
antiparticle wave functions in the outgoing bra vector back into the original
particle wave functions. The vanishing of $F^{(k)}$ is a crossing relation
which is certainly sufficient for the validity of (\ref{M}), but it does not
have the expected standard form which would result if we omit the emulation
subscript (in which case one obtains the vacuum to n-particle matrixelement of
$B)$. This is not allowed in the presence of interactions. In the following we
will show that for special ordered $\theta$-configurations the general
crossing passes to the standard form.

First we remind the reader how this was achieved in the integrable case
\cite{BFKZ} when the matrix-elements $\left\langle 0\left\vert B\right\vert
\theta_{1},..\theta_{n}\right\rangle $ are meromorphic functions. In that case
there exists, besides the degeneracy under statistics exchange of $\theta s,$
also the possibility of a \textit{nontrivial exchange via analytic
continuation}. In that case an analytic interchange of adjacent $\theta$
produces an $S(\theta_{i}-\theta_{i+1})$ factor, where $S$ is the scattering
function of the model (the two-particle S-matrix from which all higher elastic
S-matrices are given in terms of a product formula) \cite{BFKZ}. For general
permutations one obtains a representation of the permutation group which is
generated by transpositions. The steps which led to the result can be
summarized as follows:

\begin{enumerate}
\item Use the statistics degeneracy to fix a natural order so that the
"faster" particles (bigger $\theta$) are to the left of the smaller
$\theta_{1}>...>\theta_{n}$, so that in the backward extension of the velocity
lines there was no crossing of velocity lines. Identify the analytic
matrix-element in the natural order with the incoming configuration%
\begin{equation}
\left\langle 0\left\vert B\right\vert \theta_{1},..\theta_{n}\right\rangle
=\left\langle 0\left\vert B\right\vert \theta_{1},..\theta_{n}\right\rangle
_{in},~~
\end{equation}
Any other order is then determined by the analytic exchange rules in terms of
a \textit{grazing shot S-matrix~}$\mathit{S}_{gs}$%
\begin{align}
\left\langle 0\left\vert B\right\vert \theta_{1},..\theta_{n}\right\rangle  &
=S_{gs}\left\langle 0\left\vert B\right\vert \theta_{2},.\theta_{1}%
,\theta_{k+1}..\theta_{n}\right\rangle _{in},~S_{gs}=%
{\displaystyle\prod\limits_{l=2}^{k}}
S(\theta_{l}-\theta_{1})~\label{gr}\\
\theta_{2}  &  >..>\theta_{1}>\theta_{k+1}..>\theta_{n}\nonumber
\end{align}

\item The analytic exchange relation can be encoded into algebraic commutation
relations of the Zamolodchikov-Faddeev (Z-F) type%
\begin{align}
&  Z(\theta)Z^{\ast}(\theta^{\prime})=\delta(\theta-\theta^{\prime}%
)+S(\theta-\theta^{\prime}+i\pi)Z(\theta^{\prime})Z(\theta)\\
&  Z^{\ast}(\theta)Z^{\ast}(\theta^{\prime})=S(\theta-\theta^{\prime})Z^{\ast
}(\theta^{\prime})Z^{\ast}(\theta)\nonumber\\
&  Z^{\ast}(\theta_{1})..Z^{\ast}(\theta_{n})\left\vert 0\right\rangle
=\left\vert \theta_{1},..\theta_{n}\right\rangle _{in},~\theta_{1}%
>...>\theta_{n}%
\end{align}
where the last line contains the identification with the incoming particles.

\item The Z-F operators are the Fourier components of generating operators of
the interacting wedge-localized algebra \cite{S1}\cite{S2}\cite{Lech}%
\begin{equation}
A_{in}(f_{\mathcal{A}(W)}=\int_{\partial C}Z^{\ast}(\theta)e^{ip(\theta)x}%
\hat{f}(\theta)d\theta,~C=(0,i\pi)~strip,~Z(\theta)=Z^{\ast}(\theta+i\pi)
\end{equation}
where $\hat{f}(\theta)$ is the mass-shell restriction of the Fourier transform
of $f,~suppf\in W.$
\end{enumerate}

The consistency of the algebraic structure with wedge-localization and the
proven nontriviality of the intersection of double cone algebras, defined as
the intersection of two wedge-localized algebras, secure the consistency of
the analytic assumption as part of the existence a QFT whose S-matrix is the
given scattering function.

The construction has an analog for non-integrable models. The main
complication results from the \textit{presence of all inelastic threshold
singularities of multiparticle scattering} in the "analytic $\theta
$-commutation". This leads to a path-dependence for $\theta$-commutations i.e.
the analytic structure cannot be anymore subsumed into the algebraic structure
of a representation of the permutation group. For the special case (\ref{gr})
the shortest path for getting the $\theta s$ into the natural order
corresponds to the commutation of the $\theta_{1}$ with the k-1 cluster
$\theta_{2},..\theta_{k}.$ So the first question is whether there exists an
analog of the grazing shot S-matrix in the general case. For this purpose it
is helpful to rewrite the above integrable $S_{gs}$ into an expression which
only involves the full S-matrices. It is clear that
\begin{equation}
S_{gr}(\theta_{1};~\theta_{1},..\theta_{k})=S(\theta_{2},..\theta_{k})^{\ast
}S(\theta_{1},...\theta_{k})
\end{equation}
with $S$ being the full S-matrices of k respectively k-1 particles does the
job. In case the two-particle scattering matrix is not just a scattering
function but rather a matrix of scattering functions, one has to use the
Yang-Baxter relation in order to cancel all interactions \textit{within} the
k-1 cluster $\theta_{2},..\theta_{k}$; the remainder describes a "grazing
shot" of $\theta_{1}$ on the $\theta_{2},..\theta_{k}$ cluster. In this form
the grazing shot idea permits an adaptation to the general case%

\begin{align}
S_{gs}^{(m,n)}(\chi|\theta_{1};\theta)  &  \equiv\sum_{l}\int..\int
d\vartheta_{1}..d\vartheta_{m}\left\langle \chi_{1}..\chi_{m}|S^{\ast
}|\vartheta_{1},.\vartheta_{l}\right\rangle \cdot\label{gs}\\
&  \cdot\left\langle \theta_{1},\vartheta_{1},.\vartheta_{l}|S|\theta
_{1},\theta_{2},..\theta_{k}\right\rangle \nonumber
\end{align}
In this case the $\chi$ represents the $\chi=\chi_{1},..\chi_{m}$ component of
a scattering process in which the grazing shot "bullet" $\theta_{1}$ impinges
on a k-1 particle $\theta$-cluster consisting of s $\theta_{2},..\theta_{k}$
particles. Here the sum extends over all intermediate particles with
energetically accessible thresholds, i.e. the number of intermediate open
$l$-channels increase with the initial energy. The matrix elements of the
creation part of an emulat sandwiched between two multi-particle states can
directly be written in terms of the grazing shot S-matrix as%

\begin{equation}
_{in}\left\langle \chi_{1},.\chi_{m}\right\vert Z^{\ast}(\theta)_{\mathcal{A}%
(W)})\left\vert \theta_{1};\theta_{2}.\theta_{n}\right\rangle _{in}%
=S_{gs}^{(m,n)}(\chi,\theta_{1};\theta) \label{bi}%
\end{equation}
A similar formula holds for the annihilation part. Once the annihilation
operator has been commuted through to its natural position, it annihilated the
next particle on the right and contributes a delta contraction. This procedure
may be interpreted as a generalization of Wick ordering to interacting emulats.

The general grazing shot S-matrix (\ref{gs}) is the only expression which (a)
reduces to the integrable grazing shot S-matrix and (b) fulfills the
requirement that the commutation of $\theta_{1}$ with a $\theta_{2}%
,..\theta_{k}$ cluster can be expressed in terms of S-matrices only. As
mentioned this requirement has its origin in the fact that the only way, in
which the interaction enters into the theory of modular wedge localization, is
through the S-matrix. Under the assumptions (a) and (b) the commutation
formula of an emulate $\left(  A(f\right)  _{\mathcal{A}(W)}$ (or its
Wick-ordered extension) with a cluster of particles is unique and the
resulting formula may be used to evaluate the left hand side of the KMS
relation (\ref{M}) in terms of vacuum to multi-particle matrix-elements of
$B.~$The resulting formula is consistent with the standard form of the
crossing identity%

\begin{align}
&  \left\langle 0\left\vert B\right\vert \theta_{1},.\theta_{k},\theta
_{k+1}.,\theta_{n}\right\rangle _{in}=~_{out}\left\langle \bar{\theta}%
_{k+1},.,\bar{\theta}_{n}\left\vert U(\Lambda_{W_{(0.1)}}(\pi i))B\right\vert
\theta_{1},.,\theta_{k}\right\rangle _{in}\label{cro}\\
&  B\in\mathcal{A(O}),~\mathcal{O}\subseteq W_{(0,1)},~\bar{\theta
}=antiparticle~of\text{ }\theta,~~\theta_{1}>..>\theta_{n}\nonumber
\end{align}
only in case of the natural order. Any different order between the two
clusters will correspond to a different, much more complicated left hand side
which will contain contributions from grazing shot S-matrices to arbitrary
high particle number. Whereas for the partitioning of n-particle states into
two clusters the natural order can always be maintained; in case we start from
a general n-k to k formfactor, the relative ordering between out and in
$\theta$ has to be imposed in order to maintain the simple form of crossing.
Only in that case the crossing identity retains its simple form without
modification from the grazing shot S-matrix. For the special case of crossing
just one particle it reads%

\begin{equation}
_{out}\left\langle \theta_{k+1},.,\theta_{n}\left\vert B\right\vert \theta
_{1},.,\theta_{k}\right\rangle _{in}=~_{out}\left\langle \bar{\theta}_{k}%
+i\pi,\theta_{k+1}..,\theta_{n}\left\vert B\right\vert \theta_{1}%
,.,\theta_{k-1}\right\rangle _{in}%
\end{equation}
if the $\theta_{k}$ is bigger than the outgoing $\theta s.$

In the application of the Haag-Ruelle scattering theory to the derivation of
the LSZ reduction formalism \cite{B-S} there are threshold modifications from
overlapping wave functions which wreck the strong approach of the asymptotes
\cite{Buch} in the limit of large times and thus invalidate the LSZ reduction
formalism. We believe that they correspond to the opening of threshold in the
grazing shot S-matrix which enters in the algebraized analytic changes of the
natural order in the presence of overlapping wave functions.

The important new message is that the issue of the \textit{general form of the
crossing relation together with the computation of the left hand side of the
KMS identity (without the ordering restriction) is inexorably linked with a
new constructive aspect of the action of emulats on particle states} in which
the interaction enters in form of the grazing shot S-matrix. The latter
couples particle cluster (of those particles through which the emulat has been
commuted) to all sectors to which the superselection rules permits such
couplings. In other words the analytic exchange of $\theta s$ associated with
the emulate with those $\theta s$ which correspond to a cluster of particle in
the incoming state leads to a perfect realization of an on-shell version of
"Murphy's law": \textit{everything which is not forbidden to couple (subject
to the validity of the superselection rules) actually does couple}.

In off-shell QFT this is of course well-known, but on-shell (in the sense of
formfactors) this is new and somewhat surprising; it is both a blessing and a
curse. In the integrable case it leads to a representation of the permutation
group \cite{Lech} and the possibility to construct wedge generators for given
scattering function by "deformations" of free fields \cite{Le}; whereas in
general the analytic exchange is path-dependent (reflecting the influence of
the inelastic threshold cuts) and the generators require the application of
the much more complicated emulations. In fact the general situation resembles
vaguely that of a d=1+2 Wightman theory with braid group statistics \cite{B-M}
for which the Bargman-Wightman-Hall analyticity domain \cite{St-Wi} is not
schlicht but contains cuts, and the path-dependent algebraic commutation
represents the action of the infinite braid group.

The analytic $\theta$-exchanges was the crucial idea which led the authors in
\cite{BFKZ} to formulate their bootstrap-formfactor project for factorizing
models. In that case the analytic transposition of two adjacent $\theta$ can
be encoded into the algebraic Zamolodchikov-Faddeev commutation rules. In the
general case there is no transposition rule which leads to a representation of
the permutation group, rather the situation becomes analogous to the braid
group structure in which a right-left distinction is not sufficient, one must
spell out the path which led to final right-left configuration (i.e. analog of
the behind/in front move leading to braiding). We would like to think of the
process of emulation as being the analog of the functorial construction of
free field algebras from the application of the second quantization functor to
Wigner's representation theoretical construction of particles%
\[
functorial~relation\overset{interaction}{\longrightarrow}~emulation
\]
According to Mund's theorem it is impossible to maintain a functorial relation
in the presence of interactions; it has to be replaced by a bijection of
particles and their free fields into their interacting emulats.

In this way the crossing property becomes an integral part of a new
nonperturbative construction of a QFT whose first step is the construction of
wedge generators. As in the failed S-matrix bootstrap project, it is an
essential part of a new constructive program which in addition to the S-matrix
uses on-shell formfactors\footnote{In fact the matrixelements of the S-matrix
represent the formfacto of the identity operator.}. There are two unsolved
problems with this setting in the non-integrable case

\begin{itemize}
\item Show that the action of the emulates in terms of the grazing shot
S-matrices leads to wedge duality%
\begin{equation}
\left\langle \psi\left\vert \left[  J\left(  A_{in}(\hat{f})\right)
_{\mathcal{A}(W)}J,\left(  A_{in}(\hat{g})\right)  _{\mathcal{A}(W)}\right]
\right\vert \varphi\right\rangle =0,~J=S_{scat}J_{in} \label{dual}%
\end{equation}
where $\left\vert \psi\right\rangle ~$and~$\left\vert \varphi\right\rangle $
are multi-particle states and the input is an S-matrix which fulfills the
crossing property. This is a structural problem.

\item Show that the inductive use of the wedge duality starting with a lowest
order input for the S-matrix being the lowest nonvanishing mass-shell
restriction of the scattering amplitude and computing from this and the
validity of (\ref{dual}) the lowest order formfactor and afterwards the next
order S-matrix and so on. such an iteration resembles vaguely the
Epstein-Glaser iteration based on the recursive implementation of causal
locality. The divergence of perturbative series based on singular field has no
direct bearing on such an on-shell perturbation.
\end{itemize}

The main difference to the old S-matrix program is that it contains much more
structure. Without having formfactors in addition to the S-matrix and a
relation (\ref{dual}) which contains both, it is not possible to have a
constructive iteration. The hope is of course that the iteration converges,
which is known \textit{not} to be true in the case of Epstein-Glaser
perturbation which deals with singular fields.

One note of caution. The use of bilinear forms (\ref{bi}) does not mean that
the emulates $A(\hat{f})_{\mathcal{A}(W)}$ have \textit{general} n-particle
states in their domain. This is only the case for integrable models
\cite{BBS}. The emulated operators as well as their action as operators on
multiparticle states are generally only well-defined on W-localized particle
states. In computation of norms one should understand these problems in terms
of properties of the unboundedness properties of the grazing shot S-matrix.
For the check of wedge duality it is important that the operators as well as
the states are W-localized. One expects however that in formal computations
involving only matrix elements (bilinear forms instead of operators) one can
relax those requirements

The close relation of the crossing with the cyclic KMS identity underlines
again that crossing in the sense of particle theory has nothing to do with
Veneziano crossing and ST.

The present formalism replaces the old S-matrix attempts (the S-matrix
bootstrap, the dual model and string theory). It constitutes a formulation of
QFT in terms of on-shell quantities only. But it is merely the first step in a
future classification and construction setting (existence and controlled
approximations). As in the integrable case where all these steps have been
carried out \cite{Lech}, one still needs to show that the double cone
intersections of wedge-localized algebras are nontrivial.

The main message of this section is that the failure of the previous S-matrix
projects and in particular the deconstruction of string theory based on
conceptual misunderstandings about causal localization does not leave one
empty-handed. Rather one encounters a completely new window into local quantum
physics on the ruins of the old project. In this, and only in this sense, the
last 5 decades do not only constitute a loss with respect to foundational
aspects of local quantum physics; the resolution of a deep misunderstandings,
more than any so-called revolution, could be the seed of deep progress.

\section{Resum\'{e} and concluding remarks}

The main point of the present work was to explain why Mandelstam's important
project of a mass-shell based top-to-bottom approach took a wrong turn when he
mistakenly accepted Veneziano's dual model crossing as a description of the
particle physics crossing. As a result Mandelstam's farsightedness concerning
the importance of S-matrix-based on-shell projects in particle physics took a
wrong directiony as a result of his belief that this can be accomplished by
the dual model and ST.

Our derivation of crossing identities for particle formfactors was based on
the use of two important concepts which both follow from modular localization:
\textit{interacting emulats of Wick products of wedge-localized free fields}
which describe the particle content of a QFT, and their \textit{use in the KMS
identity} of modular localization. We argued that this new construction should
be interpreted as an extension of Wigner's 1939 first intrinsic
(quantization-free) construction of interaction-free local quantum
physics\footnote{The value of historical cohesion in particle physics cannot
be overestimated in times of self-proclaimed revolutions.} in terms of
positive energy representation of the Poincar\'{e} group combined with Weyl's
CCR (or its CAR counterpart) "second quantization" functor. In
contradistinction to Schr\"{o}dinger's QM, the functorial relation between
localized subspaces of wave functions and local operator subalgebras breaks
down in the relativistic case and has to be replaced by the much weaker
connection between incoming and interacting wedge algebras presented in the
previous section which replaces the functorial relation. The correct particle
crossing was shown to be an important side result of this new construction.

We showed that the dual model results from a crossing identity of conformal
correlation, using an argument which can be traced back to work by Mack
\cite{Mack}. This is sufficient to show that what Veneziano accomplished in
his construction of the dual model has no conceptual relation with the
particle crossing seen in Feynman diagrams and used as a defining property by
the protagonists of the S-matrix bootstrap as well as in Mandelstam's proposal
of a "double spectral representation". However we also added the independent
presentation of a recent in order to make our case iron-clad in order to
prevent ST to escape through conceptual back-doors or prevent them from
following their favorite strategy which Feynman characterized as "string
theorists don't make predictions, they rather make excuses".

Our proof of the particle crossing relation also confirms what (since the time
of the S-matrix bootstrap and the formulation of Mandelstam's on-shell
project) was always suspected, namely that particle crossing plays an
important role in any on-shell top-to-bottom nonperturbative construction
within particle theory. Fields without particles as those encountered in
interacting conformal QFT may play an important role as "theoretical
laboratories" for studying certain mathematical aspects of QFT (especially if
it comes to questions of mathematical existence); but no experiment will ever
measure a nucleon field; fields are the best objects to implement localization
ideas, but they are simply too fleeting for being directly measurable. Their
non-fleeting manifestation are asymptotically stable multi-localized particle
states in theories which allow a complete asymptotic particle interpretation.
In this respect the underlying philosophy in this paper is very different from
that of Wald \cite{Wald}; the ostentatious absence of Wigner's concept of
particles in curved spacetime is no reason for giving up looking for
non-fleeting entities with stable n-fold excitations of a reference state
which replaces the Poincar\'{e} invariant vacuum which are asymptotically
related with fields. 

The framework of QFT used in this paper is radically different from
quantization approaches and can be viewed as a synthesis of two hitherto
antagonistic settings: Haag's LQP approach based on localized observables and
Mandelstam's (pre-Veneziano) S-matrix setting. The new framework uses the
S-matrix already at the start, its ultimate aim is to construct local
observables. Exact solutions will always be limited to integrable models,
which in QFT are necessarily two-dimensional \cite{integrable}. In
non-integrable models with a complete particle interpretation one can only
hope for mathematically controlled approximations and with some luck, the
verification of the (hopefully unique) existence of a QFT for a given crossing
symmetric S-matrix as well as controlled arbitrary precise approximations. In
view of the suspicion that the divergence of perturbative series may be
connected to the fact that correlation of such \textit{singular objects} as
fields (operator-valued distributions) are to blame e for the lack of control,
there is even a chance that a perturbation expansion for more rugged
non-singular on-shell objects as generators of wedge-algebras may converge;
the absence of ultraviolet problems was the historical reason why Heisenberg
proposed an S-matrix setting.   

Although the new setting provides an optimistic look into a better future for
particle theory, there is good reason to be less optimistic when it comes to
the computational implementation of these new ideas in the near future. The
number of physicists familiar with foundational aspects of local quantum
physics has decreased, this is particular evident in the US where the interest
in mathematical-geometrical formalisms has overtaken that in the more subtle
foundational physical aspect in which geometric properties are always blended
with subtle physical consequences of quantum localization. 

The absence of any innovative investment into foundational knowledge ("shut up
and compute") within the new globalized communities, and with metaphoric
arguments and trendy monocultures on the rise, there is not much reason for
optimism. Also the observation that interest in string theory and its
derivatives (extra dimensions, branes, TOEs) seems to be waning cannot be a
cause for optimism. The metaphoric Zeitgeist of the ST (which seems to
parallel the metaphoric nature of financial capitalism) has entered large
parts of particle theory (extra dimensions, branes, embedding of QFTs,..).
Most supporting quasiclassical arguments collapse if it comes to problems of
real quantum matter, e.g. connections between cardinality of degree of
freedoms, time-like causality and modular localization for which one needs the
full conceptual power of LQP.

Causal localization is certainly the most subtle and far reaching property of
local quantum physics; and with the birth of QFT in the aftermath of the
incompletely understood Einstein-Jordan conundrum, there was always the latent
danger that the incompletely understood sharp conceptual borderline with
respect to the Born-localized QM could eventually cause havoc. But apart from
transitory problems with the old perturbation theory (see the old
pre-renormalization textbooks by Wenzl and Heitler) as well as the
misunderstandings about vacuum polarization in connection with the so-called
\textit{ultraviolet catastrophe }\footnote{Since the derivation of
perturbation theory from the iterative implementation of the causal locality
principle and the requirement of a perturbative order-independent bound on the
short distance behavior (the Epstein-Glaser causal perturbation) it is clear
that there is no conceptual place for ultraviolet divergencies (a principle
can either be implemented or not).}, one was able to find consistent recipes
for the new covariant perturbation \textit{without fully understanding the
relation between thermal and vacuum polarization aspects with causal
localization}.

The invocation of covariance which was the turning point in perturbation
theory (ascribed to Tomonaga) is however of not much avail if it comes to
subvolume fluctuation problems as in the Einstein-Jordan conundrum of the
cosmological constant problem; here a more direct understanding of
thermalization through causal localization is asked for. In special cases,
where modular localization leads to geometric Hamiltonians as in the case of
the Unruh effect, this was understood in terms of concrete calculations (and
only by a much smaller number of physicists as a result of structural
arguments \cite{Sewell}). The insight that important properties of particle
theory as the crossing properties of on-shell objects (scattering amplitudes,
formfactors), and a new settings for \textit{intrinsic} on-shell constructions
of particle theories (in the spirit of Mandelstam's post dispersion theory
S-matrix attempts i.e. no use Lagrangian or functional quantization) is of a
more recent vintage.

Far from leading only to a critical evaluation of what has been done in the
past, the purpose of the new approach based on modular localization also
includes to extend renormalized perturbation theory involving higher spin
fields by lowering their short-distance dimensions by using string-localized
"potentials" in Hilbert space instead of pointlike "field strengths" in Krein
spaces (BRST-formalism). In the zero mass case this includes also the project
of to understand what hitherto has been swept under the rug by referring to
"non-perturbative long distance infrared singularities" of Yang-Mills
couplings in terms of overlooked long-distance changes caused by semiinfinite
string-localized "potentials". This also effects the conceptual position of
the Higgs phenomenon and of asymptotic freedom\footnote{The remark refers to
conceptual strengthening resulting from repatriating "orphaned" beta functions
to their conceptual home of perturbatively derived Callan-Symanzik equations
for massive Y-M interactions in the string-localized setting (the massive
pointlike BRST correlations have no massless limit, the origin of the standard
infrared problems). }.

Ever since its birth in the aftermath of the 1925 Einstein-Jordan conundrum
the insufficient understanding of the physical consequences of quantum causal
localization was threatening to limit the full exploration of QFT beyond
perturbation theory. Only in the aftermath of the successful particle
adaptation of the Kramer-Kronig dispersion relation \cite{causal} when
physicists started to become interested in non-perturbative on-shell aspects
the lack of understanding hit particle theory with full vengeance. As
described in detail in section 2 its first casualty was the dual model, or to
be more precise the \textit{dual model as a description of particle crossing}.
The correct particle crossing follows from the KMS property of
wedge-localization; the same KMS property which, as a thermal epiphenomenon of
modular localization, is at the bottom of the Unruh effect and similar
situations in curved space time (in which the imagined causal horizon is
replaced by a less fleeting event horizon).

It will take a very long time to dispose of all the conceptual clutter created
not only by ST, but also by its derivatives which later became disconnected
from ST. This includes the ides of \textit{embedding of lower dimensional QFTs
in higher ones} and its inverse namely \textit{a QFT version of the quantum
mechanical Klein-Kaluza dimensional reduction}; but it also extends to
Maldacena's belief that the mathematical AdF-CFT correspondence is capable of
relating two physical theories and to the closely related incorrect idea that
the restriction of a QFT to a "brane" describes a physical QFT which fulfills
timelike causality\footnote{In \cite{Mack} it is shown that the thinning out
of degrees of freedom in holographic projections onto null-surfaces does not
happen in projections to branes.}. Such misunderstandings of local quantum
physics are often supported by "massaging" Lagrangians or using quasiclassical
approximations (which contain no information about degrees of freedom their
connection with quantum localization) which are contradicting the holistic
consequences of causal localization for local observables and their
correlations. The best way out of these confusions which arose from the
incorrect idea that quantization preserves classical/semiclassical results
about embeddings or K-K dimensional reductions, is to follow the guidance from
the modular localization theory.

Leaving string theory aside, we permitted ourselves the fascinating historical
dream of imagining a changed path of history in which the Einstein-Jordan
conundrum was solved at the time of birth of QFT. As it is well known,
Einstein had deep-rooted misgivings about Born's assignment of probability to
individual quantum mechanical events. Of course Born did this not out of the
blue, but as a consequence of an interpretative necessity in order to relate
the (Born approximation of) scattering amplitudes (nowadays the "cross
section") to the statistics of scattering events caused by sending a beam onto
a target\footnote{The application to Schr\"{o}dinger wave functions
(apparently attributed to Pauli) was later added as a footnote to Born's
famous paper.}.

Just image that Einstein would have been aware of that intrinsic thermal KMS
aspect of localization which is implicit in Jordan's QFT contribution to the
E-J conundrum. The probabilistic aspect of statistical ensembles (different
from the assignment of probabilities to single events observed on an
individual systems) was Einstein's centerpiece of his theoretical fluctuation
arguments concerning the corpuscular nature of light which got Jordan into the
E-J conundrum dispute. The clear recognition that subvolume-reduced  QFT which
unlike QM represents an \textit{ensemble of operators} belonging to the same
localized algebra and that the reduced vacuum defines a special kind of
thermal statistical analog (or even isomorphic) to a global heat-bath
statistical mechanics system could have changed history. Certainly Jordan
would have appreciated to receive Einstein's full support in his struggle
against the resistance of his coauthors Born and Heisenberg to concede a
separate section to his "wave quantization" in the Dreim\"{a}nnerarbeit (which
they grudgingly did even without Einstein's support).

This viewpoint of dealing only with ensembles was fully realized in Haag's
1957 theory of \textit{local nets of operator algebras,} in fact it is the
central issue which distinguished this approach from other settings as
Wightman's formulation.

It is Born's quantum mechanical assignment of probabilities to individual
objects which Einstein objected to, and far from criticizing him for his
stubbornness on this point, \textit{we should admire him for his steadfastness
coming from his philosophical realism}. It is the more fundamental QFT as
compared to QM which brings back to QT some of Einstein's realism by placing
various counter-intuitive effects into a more realistic light by relating the
probabilistic ensemble property of QFT as coming directly from the quantum
adaptation of the Faraday-Maxwell "action at the neighborhood principle"
instead of adding them by fiat as the Born probability in QM.

Not only the many counter-intuitive consequences of QM would have appeared in
a different (more Einstein-friendly) philosophical light, but the entire
evolution of QFT might have taken a different direction\footnote{Being the
more fundamental theory, the limitation in forcing QFT to dance to the tune of
a quantization parallelism would have been perceived much earlier. The
"ensemble" point of view in terms of extracting everything from localized
subalgebras instead from individual localized operators matches perfectly the
way an experimenter does particle physics without knowing (or even wanting to
know) the inner workings of his measuring devices.} (and certainly ST would
have remained without support). Also papers on superluminal propagation as,
they appeared in the past (and still appear almost every year), would never
have passed the editorial hurdle of PRL\footnote{To the exculpation of PRL it
should also be mentioned that immediately afterwards they published a correct
presentation of Fermi's argument that subluminal phenomena are absent in QED
\cite{Bu-Yn}.} \cite{PRL}.

It is a bit unfair to blame \textit{only} string theorists for the present
schism in particle physics. Those few individuals who always had the knowledge
about QFT's deeper conceptual layers did not leave their ivory
tower\footnote{After Res Jost's forceful critique of some of the claims in the
context of the the bootstrap approach \cite{Jost} the Streitkultur of the
Pauli era came to an end.} (which perhaps would have meant a temporary
interruption of their own work and a continuation of the old Streitkultur
which kept particle physics healthy for several decades starting from the 50s
and ending in the 70s) share part of the blame. The present work comes too
late to have an effect; but at least there is the small consolation of having tried.

In no country the impact of ST and its derivatives has been as disastrous as
in Germany, the country in which QT started. Take as an example the
theoretical physics at the university of Hamburg which after its foundation in
1920 adorned itself with an illustrative continuous line of names: Lenz,
Pauli, Jordan, Lehmann, Haag and Fredenhagen, but finally could not resist the
outside pressure of laboratories of "big science" which, in the tradition of
royal courts and their court jesters compete to have their own local string theorists.

Fact is that foundational research on QFT in a country where it began in 1925
and continuously evolved up to the present may not have a place of location in
the foreseeable future. It is a small consolation that the situation in
Austria, Italy and the UK is not as bad; for the time-being the swan-song of a
disappearing foundational research in QFT only threatens to affects the
country where it all begun.

Concluding this resum\'{e} of foundational aspects of QFT including a
scientific critique of ST, it may be helpful to the reader to comment on some
sociologal arguments which defenders of ST often use. To be concrete, I
explicitly refer to a recent published defense by Duff within a project
{\normalsize \textquotedblleft Forty Years Of String Theory: Reflecting on the
Foundations\textquotedblright\ \cite{Duff} as well as on an article by Smolin
\cite{Smolin}, who in Duff's article serves as his punching bag only to end in
his own article as a fervent subscriber to ST's most bizarre solution of the
background independence problem in form of the "multiverse"\footnote{According
to the motto: if the tail of the dog does not wiggle, then perhanps the tail's
dog can be made to wiggle.}}.

Without wanting to defend Nancy Cartwright's somewhat extreme points of view
on  emerging unification\footnote{I think what she is probably criticising is
an enforced unification which does not emerge from the natural flow of
improved insights into nature.} and the strife for a "Weltformel" or a TOE (a
unique theory of everything) against Duff's critique, it should be said that
at least its inverse, namely \textit{to conclude from the existence of a
unique or nearly unique realization} \textit{of some idea} \textit{that it
must have foundational physical significance}, is (hopefully) not acceptable
to me but also to him. But it appears that string theorists do precisely this
when they conclude from the (nearly, up to M-theoretic modifications) unique
possibility to represent a unitary positive energy representation of the
Poincar\'{e} group on what they call the "target space" of a nonrational
chiral sigma model (in more old-fashioned terminology the only solution of the
Majorana project).

Instead of trying to understand why nonrational chiral theories with their
continuously many superselected charge sectors\footnote{Different from higher
dimensional observable algebras which, at least in theories with compactly
localizable superselected charges, only permit extensions to charged algebras
with compact internal group symmetries nonrational chiral QFTs are the (little
explored) breeding ground for representations of noncompact inner symmetries.}
allow representations of noncompact "inner symmetry" groups, string theorists
insist to solve the problem by identifying our living spacetime with the
noncompact internal symmetry space of a nonrational sigma model. Its
continuously superselected charge structure constitutes the only known cases
for realizing noncompact inner symmetries on the oscillator content of
compactified chiral conformal QFT; the best (and presently only known)
illustration for the realization of a positive energy representation of the
Poincar\'{e} group is the superstring representation on a 10-component abelian
chiral current model.

Whereas it is certainly true that one can do this on the "target" space of a
(mildly modified) chiral sigma model associated to abelian currents in d=10
spacetime dimensions, the important question is really \textit{what to make of
such constructions}. Should one interpret this property abstracted from a
certain way of dealing with the charge spectrum of multi-component current
models as defining a foundational spacetime theory (the dog's tail wiggling
with the dog)? Are we living in a dimensionally reduced inner symmetry space
of a (modified) nonrational conformal QFT ? Apparently Duff \cite{Duff}
believes that we are.

One has all reasons for being somewhat surprised about the near uniqueness (up
to a finite number of M-theoretic changes) of the 10 dimensional superstring
representation of the Poincar\'{e} group on the target space of a nonrational
sigma model, but there is no conceptual justification to interpret the rarity
of such an occurrence (probably related to the fact that non-rational chiral
theories have not been the subject of systematic studies) as the harbinger of
a new foundational insight into spacetime. Would particle physics in the last
5 decades have developed in a different way if the answer to this question
would have been that there is \textit{either no or infinitely many
representations} ?

Attributing to this observation the role of a key to the understanding of the
universe is not much different than the ontological role \textit{attributed to
the number 42 as an answer to the ultimate question about "Life, the Universe,
and Everything"} in Douglas Adam's scientific fiction comedy "the hitchhiker's
guide through the galaxy".

A closely related remark mentioned in section 2 is that, whereas the
principles of classical field theory admit dynamical covariant quantum
mechanical variables $X_{\mu}(\tau)$ which parametrize world-lines in any
dimension, there are \textit{no covariant quantum position operators which
correspond to these classical objects}; rather the move with changing $\tau$
takes place in an inner space "over" the localization point (where the spin
components are pictured) where localization has no relevance. This is a
special case of the general impossibility of quantum field theoretical
embeddings and K-K dimensional reductions. In any QFT with $d>1+2$ indices of
covariant fields can only refer to compact internal symmetry groups or
represent spacetime spinor/tensor indices corresponding to the spacetime
living space dimensions of the fields. Classical fields on the other hand
\textit{can also carry indices} on which noncompact groups act, but
quantization does not lead to quantum analogs. There is no law which says that
the more foundational QFT has to dance to a quantization parallelism.

As there are many classical Lagrangians which cannot be quantized, there are
also many QFT which possess no Lagrangians. The best illustration of the
latter state of affairs is given by most d=1+1 models which have been
constructed with the formfactor-bootstrap model and whose existence was
secured by Lechner's application of the \textit{modular nuclearity} idea about
the cardinality of degrees of freedom \cite{ABL}. The abundance of
non-Lagrangian QFTs in d=1+1 can be understood in terms of the larger
cardinality of crossing symmetric elastic scattering amplitudes as compared to
renormalizable Lagrangian couplings.

The sociological criticism contained in some well-known books and articles by
Woit, Smolin and others is of no help in this context. Even though this
critique may have been well intentioned, string theorists and their
adversaries began to live in symbiosis with their opponents. Sociological
critique does not penetrate the thick layer of misunderstandings around the
subtlest of all principles: quantum causal locality. It rather runs the danger
of becoming part of the scientific entertainment industry in an unfortunate
mutual dependence between supporters and "opponents".

To many particle physicists string theory appeared as the most surreal if not
bizarre subject they ever met. Hopefully this article can convince readers
that its \textit{surreal nature is not the result of a computational mistake}
nor an \textit{easy to spot} conceptual misunderstanding. The present paper
exposes some of the scientific causes of these surreal feelings. The
sophisticated nature of the conceptual mistake also means that the string
theory research during 5 decades does not amount to a total loss of time,
since the derailment of a foundational subject as the Mandelstam on-shell
top-to-bottom project is a strong motivations for a second startup using the
new property of $S_{scat}$ as a relative modular invariant of
wedge-localization as formulated in the present work. Without the recent
concept of modular localization it also would not have been possible to find a
new string-localized setting for higher spin fields which lowers their short
distance scaling dimension to the $s$-independent value $d_{s.d.}=1$ which
opens the gates for higher spin renormalized perturbation theory. It was the
critique of previous S-matrix based attempts (in particular ST) which played
an important role in obtaining these new insights and in this way led to a
golden opportunity for future progress.

To feel the depth of the crisis into which large parts of particle theory has
fallen, it is interesting to present a quotation from Einstein's talk in the
honor of Planck \cite{Quote}.

\textit{In the temple of science are many mansions, and various indeed are
they who dwell therein and the motives that have led them thither. Many take
to science out of a joyful sense of superior intellectual power; science is
their own special sport to which they look for vivid experience and the
satisfaction of ambition; many others are found in the temple who have offered
the product of their brains on this altar for purely utilitarian purposes.
Were an angel of the Lord to come and drive all these people belonging to
these two categories out of the temple, the assemblage would be seriously
depleted, but there would still be some men, of present and past times, left
inside. Our Planck is one of them, and that is why we love him. ...}

But where has Einstein's \textit{Angel of the Lord}, the protector of the
temple of science, gone in the times of string theory and its derivatives?
Reading these lines and comparing them with the content of \cite{Duff} as well
as that of his opponents, one cannot help to sense how similar the present
Zeitgeist of particle theory has become with that of the financial investment
markets. Nowhere is this better reflected than in the alternative to the Nobel
physics prize created by the Russian oligarch Yuri Milner. His in principle
noble decision to spend part of his gain from financial transactions has been
taken in the belief that by awarding the prize to protagonists of ST and
related subjects of strong sociological impact ("many people cannot err") on a
large part of the particle physics community one can correct what in his view
has been unduly overlooked by the Nobel committee. None of the individuals
with a foundational knowledge of local quantum physics of the 60s would have
proposed a dual model or ST; the incomplete knowledge or lack of knowledge
about the relation of locality with the particle crossing property was the
prerequisite for being able to do ST. Only this incorrect idea made it
possible to claim that the duality of a meromorphic function constructed by
using mathematical properties of the Euler beta-function may be related with
the particle crossing property of a an elastic scattering amplitude. 

A second chance was lost when Fubini et al. \cite{Fub} used ideas from chiral
conformal QFT for its construction and they failed to see/state that what they
identified with the particle poles were really the anomalous scale dimensions
of the composites \cite{Mack} in conformal global operator product expansions;
after this missed chance ST became the victim of this admittedly confusing
picture puzzle. Milner's prize will remind future physicists in a post ST era
about those parts of particle theory which got lost in the maelstrom of the ST
Zeitgeist, it is the first award which is given for making the best use out of
one' lack of knowledge without even being aware after it happened. Such a
prize highlights the deep schism within the actual particle physics community.

In the present paper we showed how Mandelstam's on-shell top-to-bottom idea of
accessing nonperturbative particle theory can be saved in a new approach which
avoids the dual model/ST picture puzzle confusion by using modular
localization which places the $S_{scat}$-matrix (in its role as a relative
modular invariant) together with formfactors (the matrix-elements of
$S_{scat}$ are formfactors of the identity operator!) under one new
constructive roof within LQP. The new setting is deep, and since it is still
in its beginnings one can expect the appearance of many unaccustomed aspects,
but it will never lead to such bizarre consequences as those which came in the
wake of ST \cite{Mal}.

The proximity of foundational research to what is perceived as big science is
risky, only those intellectual products which find a lobby will continue and
once a large enough community has formed they are beyond critical review. What
will future historians of physics make of such misunderstood and half-baked
ideas as ST, its bizarre derivatives as "the landscape" or of Tegmark's even
more bizarre credo that every mathematically correct ideas will have a
realization in physics one of the zillions of parallel universes?

It is perhaps not an accident that this happens while we live in a time which
moves away from the ideals of enlightenment and religious fanatism is
replacing previous economic ideologies. Is the wide-spread support of
metaphoric ideas in a particle physics part of a a Zeitgeist-phenomenon ? 

In his recent defence of ST Duff should not have left out the name of Feynman
who had, probably more driven by its bizarre appearance than its conceptual
structure, pointed out that his discussions with proponents of ST convinced
him that it is the first construct in particle theory which is not defended by
arguments but rather by taking recourse to excuses.

Duff should not be so sure about counting on Weinberg, who on several
occasions speculated that string theory may still reveal itself as a
camouflaged QFT \cite{Wei}. In fact this is precisely what the present paper
showed: the acceptable observations coming from string theory is its
identification with a \textit{dynamical} infinite component
pointlike-localized wave function where, as mentioned before, "dynamical"
means that its wave function space is obtained by solving the "Majorana
project" which consists in constructing \textit{irreducible} operator algebras
which carry a discretely reducible representation of the Poincar\'{e} group
describing an infinite collection of irreducible positive energy one-particle
representation. It was pointed out in this article that the only known
solution of this "infinite component dynamical wave function project" is the
"superstring representation" abstracted from the irreducible oscillator
algebra of a d=10 component supersymmetric chiral current model (related to
the Polyakov action). Since no zero mass infinite spin representation appears
in this decomposition the wave functions are pointlike generated and the
second quantization leads to an infinite component dynamic pointlike quantum
field. Such a pointlike field is of course too singular to pass as an
operator-valued Schwartz distribution, but by projecting to finite invariant
energy subspaces one recovers the standard situation. At the time when
Majorana \cite{Maj} looked for an infinite component relativistic analog of
the $O(4,2)$ hydrogen spectrum and also afterwards\footnote{I am referring to
the Barut, Kleinert, Fronsdal ...consstructions using extensions of the
noncompact Lorenz group (see appendix in \cite{To}).} the search was limited
to group representation algebras which extend the Lorentz group. No solution
of the Majaorana project in this limited context was found and hence the
superstring representation from the d=10 Polaykov oscillators (which also
intertwine between the levels of the infinite (m,s) tower). 

Concerning most of the other names in Duff's list on whose support of ST he
counts, one should perhaps point out that most scientist have a legitimate
natural curiosity which leads them to have an unprejudiced look at
\textit{any} new idea which isn't outright foolish. But they usually do not
sacrify much time to understand the detailed mathematical/conceptual reasons
why at second glance ST appears as a mixture of high powered mathematics with
somewhat bizarre physics; they rather simply turn away from it. Nowhere in
this paper I claim that string theory suffers from a simple-to-recognize
mistake of the kind which almost every year leads to new papers on
superluminal phenomena.

What makes the critique of ST a very tricky enterprise is the strong analogy,
if not to say picture puzzle situation, generated by conserved chiral charges
and their associated quadratic anomalous dimensions on one side and particle
momenta and their quadratic mass squares on the particle side. It is this
similarity which led to the dual model and the incorrect embedding picture
underlying ST. The only credible remainder of this analogy is that, whenever
it helps to find a representation of the Poincar\'{e} group on the substrate
of chiral oscillators, the mass spectrum of this representation is a
subspectrum of anomalous dimensions of a chiral conformal QFT. Why does one
not find this statement together with a commentary in ST
papers\footnote{Actually Fubini and collaborators came quite close
\cite{Fub}.}? It is a clear indication that one is outside the setting of
particle crossing.

With respect to Witten, Duff raises another interesting point: the relation of
string theory with mathematics. String theoretic pictures have indeed been
helpful to generate proven and provable mathematical conjectures, but does
this mean (as Duff seems to suggest) that such a strong autonomous science as
particle physics should be happy in a role of a subcontractor of mathematics?
Mathematicians need not care whether they get their inspirations from (what
they conceive as) beautiful castles within flourishing landscapes of particle
physics or from its ruins; as long as something represents a fertile soil for
their imagination and makes them free to work a bit outside the standard
conjecture-theorem-proof pattern they may profit from this source.

One of the populated mathematics-physics meeting grounds since the early 70s
is geometry. I hope that the presentation of modular localization in this
essay made clear that this means something very different for mathematicians
as it does (or rather should do) for particle theorists. Whereas in areas of
classical physics, notably general relativity and classical gauge theory, the
important role of geometry cannot be denied, the connection of QM with
geometry is less tight and starts to drift apart in QFT.

The reason is that spacetime geometry in QFT never appears \textit{without}
being burdened with vacuum polarization and thermal KMS properties as
localization entropy \cite{integrable}. To give an illustration, the
Atiyah-Singer index theory has applications to QM and free QFT in external
fields, but has no place in local quantum physics when localization comes
together with vacuum polarization and thermal manifestations. Another
illustration comes from the WZWN model. A topological Lagrangian as e.g. the
topological Euclidian WZWN action has little to do with the original
Wess-Zumino Lagrangian, which at least formally complies with Lagrangian
quantization and vacuum polarization (prerequisites for the validity of a
causal perturbation theory) and as a consequence admits the applications of
perturbation theory. The WZWN topological Lagrangian serves for "baptizing" a
model in the traditional way by reading the representation theory of d=1+1
chiral currents and the construction of the associated sigma-model fields back
into a classical Lagrangian setting at the prize of a topological euclidean
extension into a third dimension; but it is precisely this topological aspect
which is in conflict with causal localization and prevents the applicability
of the standard perturbation formalism. When it comes to computations, one
resorts to the representation theoretical methods for currents and their
associated sigma-model field (exponentials of potentials of currents). The
beauty of a WZWN Lagrangian is entirely on the geometric-mathematical side; it
becomes rather useless for the study of localization and its physical
consequences as renormalized perturbation theory.

On the other hand mathematicians have undeniably a natural intrinsic interests
in such topological Lagrangians; they are not concerned with the problems of
quantum localization. With todays hindsight about the past it is clear that
the attempts to bring geometry and quantum physics together as expressed in
the Atiyah-Witten project of the 70s and its later continuation in the setting
of string theory had their greatest success in raising the level of
mathematical sophistication of physicists, rather than advancing the course of
particle physics.

The geometrical visualization in terms of Riemann surfaces of analyticity
properties of duality of chiral models in thermal states has no relation to
the "living space" (in the sense of localization) of those models but rather
correspond to the Bargman-Hall-Wightman domain of analyticity of correlation
functions. Last not least objects of string theory are not localized on
strings; the oscillator degrees of freedom which, thinking in terms of quantum
mechanical chains of oscilators, ST incorrectly envisages as spacetime string
have no bearing on spacetime localization at all, they are simply "sitting" in
an \textit{inner space} which is attached to a spacetime localization point
(there were one imagines spin components and matrices acting on them to be living).

The imitation of Feynman rules by world-sheet pictures is a \textit{metaphoric
step which is not supported by any conceptual understanding} of quantum causal
localization; it just tries to extend a helpful way to organize perturbation
theory in terms of graphical rules outside its range of validity. To place
this into an ironical historical context, one may say that St\"{u}ckelberg was
very lucky when he extended his graphical illustration of the asymptotic
rescattering structure (which follow from his macrocausal ideas about the
asymptotic one-particle structure of relativistic scattering theory) to
non-asymptotic regions and in this way arrived at the Feynman graphs before
Feynman. But this has no repetition on the level of world sheets; the
perturbative rules for the string-localized potentials in section 3 are
entirely different.

The historical episode where quantum physics and mathematics were totally on
par was not (as most people think) the discovery of QM\footnote{The Hilbert
space theory already existed and it were not the physicists in Hilbert's
G\"{o}ttingen but rather a research assistent at the technische Hochschule in
Stuttgart namely Fritz London, who used "rotations in Hilbert space" (nowadays
unitary operators) for the first time in QM (the first paper on transformation
theory).}, rather it was the parallel development in the middle of the 60s of
what physicists called "statistical mechanics of open systems" and
mathematicians "the Tomita-Takesaki modular theory of operator algebras". It
was a taking and giving on totally equal terms between the T-T modular theory
on the mathematical side and the statistical mechanics of open systems. It
prepared the ground for the modular localization theory which started a decade
later and eventually led to the deep connections between operator algebras
generated by chiral conformal fields and Vaughan Jones subfactor theory
\cite{Jones}.

This perfect meeting of mathematical and particle physics minds also finds its
expression in the (unjustly little known) Doplicher-Haag-Roberts
superselection analysis \cite{Haag}\cite{DR} leading to compact group duals in
which the Markov traces of the Vaughan Jones theory of subfactors \cite{Jones}
and the use of endomorphisms were used independently from that of subfactor
theory; after realizing that they hit the same structure in a different
context this significantly accelerated the explicit construction of chiral
conformal models including proofs of existence.

From a philosophical viewpoint the superselection theory achieved a deep
spacetime understanding of the quantum origin of Heisenberg's phenomenological
concept of SU(2) "isospin" in nuclear physics by showing that group theory is
a surprising consequence of the classification of equivalence classes of
localizable representations of observable algebras. This is almost as
surprising as the intrinsic probability which comes from modular localization
without referring to Born's quantum mechanical probability.

There is one issue in which Smolin \cite{Smol} (together with Arnsdorf),
standing on the shoulders of Rehren\footnote{It is truely admirable how, in
the face of concentrated misunderstandings, Rehren succeeds to maintain his
countenance \cite{blog}.} should be supported against Duff. These authors
pointed at a kind of conundrum between the bizarre consequences of the
string-induced Maldacena conjecture \cite{Mal} and Rehren's theorem. In a
previous paper by Rehren \cite{Reh} it was pointed out that the rigorous
correspondence, though mathematical correct, has a serious physical
shortcoming. One side of this mathematically well-defined correspondence is
always unphysical; if one starts from a physical model on the AdS side, the
CFT side will have way to many degrees of freedom many of them are not in the
initial data but rather enter the causal dependency region "sideways" as
poltergeists. In the opposite direction i.e. starting from a physical AdS
model will be too "anemic" to support nontrivial causal localization in
compact spacetime regions.

This means that although the correspondence respects local commutativity
(Einstein causality), it violates the quantum analog of what one calls
classically the causal propagation (and becomes the causal closure (or
time-slice) property  in LQP) i.e. the algebra of the causal completion of a
region $\mathcal{O\rightarrow O}^{\prime\prime}$ is larger than that
associated with $\mathcal{O~}:~\mathcal{A(O})\subsetneqq\mathcal{A(O}%
^{\prime\prime})$ as a result of additional degrees of freedom appearing from
nowhere. For the "occupants" of its causal completion (the causally closed
double cone world) this is like a "poltergeist" effect; degrees of freedom
seemingly "coming in from nowhere" (not contained in the slice data) appear in
the causal shadow and destroy the validity of timelike causality. QFT models
accessed by Lagrangian quantization do not have this physical pathology; the
\textit{time slice postulate of QFT}\footnote{In a modern setting the
principle of causal localization comprises two requirements on observables:
Einstein causality (spacelike commutativity) and Haag duality
$\mathcal{A(O)=A(O}^{\prime\prime})$ (timelike causal propagation).}
\cite{H-S} was precisely introduced in order to save from Lagrangian field
theory what should be saved in a world outside quantization (as it is needed
in the AdS-CFT problem).

The violation of this property is intimately related to the \textit{phase
space degrees of freedom} issue which led Haag and Swieca \cite{H-Sw} to their
result that, different from QM (with or without second quantization) which
leads to a finite number of states per cell in phase space, QFT as we know it
from quantization requires a compact set (later refined to "nuclear"
\cite{Haag}). This "mildly" infinite cardinality of degrees of freedom secures
the existence of heat bath temperature states for arbitrary temperatures as
well as causal propagation. It seems that this kind of insight together with
other deep pre-electronic insights got lost in the maelstrom of time and that
especially ST remained ignorant about its existence.

Ignoring the degree of freedom requirement for a moment before returning to it
later on, one can ask the question whether it is possible to slightly modify
the AdS-CFT setting, so that a modified and appropriately reformulated
Maldacena's conjecture is in harmony with Rehren's rigorous theorem. This is
precisely the question Kay and Ortiz asked \cite{Kay}. Taking their cue from
prior work on the correspondence principle of Mukohyama-Israel as well from 't
Hoofts brick-wall idea\footnote{A physical argument which leads to a
vacuum-polarization driven entropical area law which seems to be closely
related to the dependence on the thickness of the fuzzy surface $\varepsilon$
associated with the localization entropy as defined by the split property with
$\varepsilon$ being the size of the split \cite{Haag}.} \cite{'t Hooft}, these
authors start with a Hartle-Hawking-Israel like pure state on an imagined
combined matter + gravity dynamical system. They then propose to equate the
AdS side of a hypothetical conformal invariant supersymmetric Yang-Mills model
with the restriction of the H-H-I state to a matter subsystem in accordance
with Rehren's theorem.

It is conceivable that the "degrees of freedom mismatch" in Rehren's theorem,
which in the $\rightarrow$ direction leads an \textit{overpopulation} of
degrees of freedom and in $\leftarrow~$to an \textit{anemia} can be repaired
by some extension of the Kay-Ortiz scheme, but is not very plausible.
Nevertheless their results, although as stated by the authors not rigorous,
are sufficiently interesting and deserve to be taken serious by the community
around the Maldacena conjecture. Physics is one of the few science were errors
about important properties should receive no lesser attention than correct
observations. 

In illustrating the "unreasonable power of less than perfect discoveries",
Duff points convincingly at the story of antiparticles emerging from Dirac's
hole theory. Indeed some discoveries, especially in the beginning of QFT, did
not follow straight logical lines. It was not (as one would have expected)
Pascual Jordan, the discoverer of QFT and the positivistic advocate of
quantizing everything (Maxwell fields, matter fields) which permits to be
quantized, who first saw the relation between charge and antiparticles, rather
it was Dirac who distilled the suggestion of antiparticles from his (later
abandoned) hole theory. As an intrinsic property of the underlying causal
localization aspect of QFT this was shown later by Jost \cite{Jost}.

However it was Dirac's particle hole theory which gave rise to the idea of
antiparticles, even though conceptually it shouldn't. His philosophical
setting for QT was quite different from Jordan's positivism since he used wave
quantization \textit{only} for classical Maxwell waves and described massive
matter in terms of QM. It was somewhat artistic to see antiparticles in the
context of hole theory; in fact this setting was later abandoned after it
became clear that it becomes inconsistent as soon as vacuum polarization comes
into play\footnote{It was still used in the first textbooks by Wenzel and
Heitler but did not survive renormalized perturbation theory where vacuum
polarization became important.}. Dirac came around to embrace universal field
quantization in the early 50s.

As far as stressing old-fashioned virtues in particle theory, there is no
problem to agree with Duff. This also includes his refutation of a time-limit
on string theory research as expressed in the papers of Woit and Smolin. If
string theory really would be what it claims to be, namely a consistent theory
which goes beyond QFT, it certainly has the right to take as much time as it
needs to settle this problem; but the point is that isn't.

Duff forgot to tell what he considers to be the string theoretic analog of
Dirac's discovery. Also his mentioning of the Higgs mechanism and gauge theory
in connection with string theory warrants some corrective remarks. The present
day view of massive vectormesons by a Higgs symmetry breaking and the Higgs
particle playing the role of "God's particle" (giving masses to the other
particles) is what the maelstrom of time left over from a much richer past
which in the 70ies was referred to as the \textit{Schwinger-Higgs screening
mechanism} \cite{Swieca}. The Higgs model is nothing else than the
charge-screened mode of scalar electrodynamics. Whereas the quantum mechanical
Debye screening only generates a short-range effective interaction, the QFT
screening is more radical in that it affects also the particle spectrum.
Charge screening means that the integral over the charge density vanishes,
whereas a (spontaneous) symmetry breaking brings about a divergence of this
integral (as a result of its bad infrared behavior caused by the coupling of
the conserved current to the massless Goldstone boson). The problem of why the
implementation of screening in massive QED within a BRST setting needs no
Higgs particle, whereas its nonabelian massive YM counterpart the BRST
consistency requires the presence of such physical of freedom can obviously
not be answered within the BRST setting. The answer is expected to come from
the (ongoing) perturbative computations with stringlike vectorpotentials
(section 3).

In contradistinction to the metaphoric idea about the Higgs particle
generating the mass of other particles (and presumably also of itself) which
\textit{is not an intrinsic property} and therefore is not accessible to
measurements, the divergent charge of a spontaneously broken symmetry and a
vanishing charge of a Schwinger Higgs screening mechanism are physically
distinct phenomena and according to Swieca's charge screening theorem a
massive photon is inexorably connected with charge screening \cite{Swieca} and
not with spontaneous symmetry breaking.

The idea of string-localized vector potentials also re-opens the question of
alternatives to the Schwinger-Higgs screening \cite{charge}\cite{nonlocal}.
Spinor-QED has a massive counterpart \cite{L-S} (without introducing S-H
screening via an additional Higgs degree of freedom), the so-called
\textit{massive QED,} which in the pointlike formalism needs an intermediate
BRST ghost formalism in order to lower the scaling dimension of the effective
vectorpotential from 2 to 1. This problem has meanwhile been solved in the
setting of string-localized potentials \cite{Jens}. It may be interesting to
try to substitute the BRST formalism by string-localized free vector fields
which also have short distance dimension d=1 since only string-localized
potentials allow a smooth transition between the massive and the massless
case. For Yang-Mills theory there is a "perturbative theorem" that the
consistent use of the BRST formalism requires the presence of additional
physical degrees of freedom (the scalar Higgs boson). Alternatives to this
construct have never been pursued during the 40 year history of the standard
model. This adds an air of desperation to the search of the Higgs boson which
is somewhat detrimental for the credibility of a positive result, but perhaps
explains the hype around the first traces of a new particle in the expected
energy range of the Higgs particle.

String theory has (contrary to what Duff claims) led to stagnation of vital
parts of particle physics\footnote{The conceptually confused situation
actually hinders young people with brilliant computional abilities to reach
their true innovative potential.}. What does Duff (or anybody else) expect
from a theory which, unlike all other theories has no pre-history and is
already misleading in the terminology of its name? The conceptual harbinger of
QM was the semiclassical Bohr-Sommerfeld theory and that of QFT the dispute
between Einstein and Jordan \cite{E-J}. A theory which suddenly pops up from
nowhere as the dual model (tinkering with mathematics) has a good chance to
also go nowhere.

With the widespread acceptation of string theory, Einstein's epoch of viewing
theoretical physics as a process of unfolding physical principles and
concepts\footnote{His belief in the superiority of dicovering underlying
principles before experimentally verifying their consequences was so string
that in certain cases he statet that if it does not work this way the Dear
Lord "missed a chance".} seems to have come to an end. The total break with
this Einsteinian tradition would be attained with the acceptance of such ideas
as the \textit{physical realization of any consistent mathematical structure}
in one universe of a multiverse \cite{Teg}); already the presentation of the
multiverse as a solution of the background indepence problem is way off the
mark. In fact this way of thinking became a self-runner, it does not need any
more the support of string theory; it has generated its own fantasy world
(extra dimensions \cite{Lisa}, branes as physical subsysystems, Maldacena's
claim that the AdS-CFT correspondence relates to physical theories \cite{Mal} etc.)

It will be a long lasting task for the coming new generations to remove all
the metaphoric rubble around "theories of everything" in order to have a
chance to successfully confront the LHC experimental results with new ideas
going beyond the loose ends of the standard model. The immense progress one
could expect from correcting these (by no means trivial) conceptual errors
should be a consolation for the many lost decades.

\textit{Acknowledgement}: I am indebted to Bernard Kay for pointing to his
attempt to solve the Arnsdorf-Smolin conundrum. Thanks go also to Raymond
Stora for discussions about the conceptual state of Yang-Mills perturbative
renormalization theory.


\begin{thebibliography}{99}                                                                                               %


\bibitem {Du-Ja}A. Duncan and M. Janssen, \textit{Pascual Jordan's resolution
of the conundrum of the wave-particle duality of light}, arXiv:0709.3812

\bibitem {E-J}B. Schroer, \textit{The Einstein-Jordan conundrum and its
relation to ongoing foundational research in local quantum physics}, to be
published in EPJH, arXiv:1101.0569

\bibitem {Unruh}W. G. Unruh, \textit{Notes on black hole evaporation}, Phys.
Rev. \textbf{D14}, (1976) 870-892

\bibitem {Sewell}G. Sewell, Ann. Phys. \textbf{141}, (1982) \ 201

\bibitem {Ho-Wa}S. Hollands and R. M. Wald, General Relativity and Gravitation
\textbf{36}, (2004) 2595

\bibitem {Haag}R. Haag, \textit{Local Quantum Physics}, Springer 1996

\bibitem {Jo}B. Schroer, \textit{Pascual Jordan's legacy and the ongoing
research in quantum field theory}, arXiv:1010.4431

\bibitem {Mandel}S. Mandelstam, Phys. Rev. 128, (1962) 1474

\bibitem {Maj}E. Majorana, Teoria relativistica di particelle con momentum
internisico arbitrario, Nuovo Cimento 9, (1932) 335

\bibitem {To}N. N. Bogoliubov, A. Logunov, A. I. Oksak and I. T. Todorov,
\textit{General principles of quantum field theory}, Dordrecht Kluwer

\bibitem {Polch}J. Polchinski, \textit{String theory I}, Cambridge University
Press 1998

\bibitem {Schwe}S. S. Schweber, \textit{QED and the men who made it; Dyson,
Feynman, Schwinger and Tomonaga}, Princeton University Press 1994

\bibitem {Ven}G. Veneziano, Nuovo Cim. A \textbf{57}, (1968) 190

\bibitem {1957}R. Haag, \textit{Discussion of the `axioms' and the asymptotic
properties of a local field theory with composite particles }(historical
document), Eur. Phys. J. H 35, 243--253 (2010)

\bibitem {Freden}B. Schroer, \textit{Pascual Jordan's legacy and the ongoing
research in quantum field theory}, arXiv:1010.4431

\bibitem {St-Wi}R. F. Streater and A. S. Wightman, PCT Spin\&Statistics and
all that, New York, Benjamin 1964

\bibitem {E.G}H. Epstein and V. Glaser, Ann. Inst. Henri Poincare A XIX,
(1973) 211

\bibitem {S1}B. Schroer, \textit{Modular localization and the
bootstrap-formfactor program}, Nucl. Phys. \textbf{B499}, 1997, 547, hep-th/9702145.

\bibitem {S2}B. Schroer, \textit{Modular localization and the d=1+1 formfactor
program}, Annals of Physics \textbf{295}, (1999) 190

\bibitem {Lech}G. Lechner, \textit{An Existence Proof for Interacting Quantum
Field Theories with a Factorizing S-Matrix}, Commun. Mat. Phys. \textbf{227},
(2008) 821, arXiv.org/abs/math-ph/0601022

\bibitem {Le}G. Lechner, \textit{Deformations of quantum field theories and
integrable models}, arXiv:1104.1948

\bibitem {E-G}H. Epstein and V. Glaser, Ann. Inst. Henri Poincare A
\textbf{XIX}, (1973) 211

\bibitem {Ka}H. Babujian and M. Karowski, Int. J. Mod. Phys. \textbf{A1952},
(2004) 34, \ and references therein to the beginnings of the
bootstrap-formfactor program

\bibitem {integrable}B. Schroer, \textit{The foundational origin of
integrability in quantum field theory}, to be published in Foundations of
Physics, arXiv:1109.1212

\bibitem {Mack}G. Mack, \textit{D-dimensional Conformal Field Theories with
anomalous dimensions as Dual Resonance Models},\ arXiv:0907.2407

\bibitem {BEG}J. Bros, H. Epstein and V. Glaser, Com. Math. Phys. \textbf{1},
(1965) 240

\bibitem {BFKZ}H. Babujian, A.Fring, M. Karowski and A. Zapletal, Nucl. Phys.
B\textbf{ 538}, (1990) 535

\bibitem {Jost}R. Jost: TCP-Invarianz der Streumatrix und interpolierende
Felder. Helvetica Phys. Acta \textbf{36}, (1963) 77

\bibitem {Wei}S. Weinberg, \textit{What is Quantum Field Theory, and What Did
We Think It Is}?, arXiv:hep-th/9702027

\bibitem {BRZ}D. Bahns, K. Rejzner and J. Zahn, \textit{The effective theory
of strings}, arXiv:1204.6263

\bibitem {Fred}K. Fredenhagen, K. Rejzner, \textit{Local covariance and
background independence}, arXiv:1102.2376

\bibitem {Pohl}D. Bahns, J. Math. Phys. \textbf{45}, (2004) 4640

\bibitem {Lowe}D. A. Lowe, Phys. Lett. B 326, (1994) 223

\bibitem {Martinec}E. Martinec, Class. Quant. Grav. \textbf{10}, (1993) 187

\bibitem {BMT}D. Buchholz, G. Mack and I. Todorov, Nucl.Phys. B, Proc. Suppl.
\textbf{5B}, (1988) 20

\bibitem {Sta}C.P. Staskiewicz, Die lokale Struktur abelscher Stromalgebren
auf dem Kreis, Thesis at Freie Universitaet Berlin (1995)

\bibitem {Ka-Lo}Y. Kawahigashi and R. Longo, Adv. Math. \textbf{206} (2006)
729, arXiv:math/0407263

\bibitem {Weinbook}S. Weinberg, \textit{The Quantum Theory of Fields I,
}Cambridge University Press

\bibitem {charge}B. Schroer, \textit{An alternative to the gauge theory
setting}, Found. of Phys. \textbf{41}, (2011) 1543, arXiv:1012.0013

\bibitem {nonlocal}B. Schroer, \textit{Unexplored regions in QFT and the
conceptual foundations of the Standard Model}, arXiv:1010.4431

\bibitem {BDR}D. Buchholz, S. Doplicher and J. Roberts, in preparation

\bibitem {Scharf}G. Scharf, Quantum Gauge Theories, a true ghost story, John
Wiley\&Sons, INC 2001

\bibitem {Go-Lo}M. Gomes, J. H. Lowenstein, Nucl. Phys. \textbf{B45,} (1972) 252

\bibitem {Rio}J. Mund, Prog. Math. \textbf{251}, (2007) 199, arXiv:hep-th/0502014

\bibitem {Jens}J. Mund, private communication

\bibitem {future}J. Mund, B. Schroer and J. Ynvason, in preparation

\bibitem {BGL}R. Brunetti, D. Guido and R. Longo, \textit{Modular localization
and Wigner particles}, Rev. Math. Phys. \textbf{14}, (2002) 759

\bibitem {Fa-Sc}L. Fassarella and B. Schroer, \textit{Wigner particle theory
and local quantum physics}, J. Phys. A \textbf{35}, (2002) 9123-9164

\bibitem {MSY}J. Mund, B. Schroer and J. Yngvason, \textit{String-localized
quantum fields and modular localization}, CMP\textbf{\ 268} (2006) 621, math-ph/0511042

\bibitem {Bo}H-J. Borchers, \textit{On revolutionizing quantum field theory
with Tomita's modular theory}, J. Math. Phys. \textbf{41}, (2000) 8604

\bibitem {interface}B. Schroer, \textit{Studies in History and Philosophy of
Modern Physics }\textbf{41} (2010) 104--127, arXiv:0912.2874

\bibitem {W-K}R. Kaehler and H.-P. Wiesbrock, \textit{Modular theory and the
reconstruction of four-dimensional quantum field theories}, Journal of
Mathematical Physics \textbf{42}, \ (2001) 74-86

\bibitem {BBS}H. J. Borchers, D. Buchholz and B. Schroer, Commun. Math. Phys.
\textbf{219} (2001) 125

\bibitem {Mund2}J. Mund, \textit{An Algebraic Jost-Schroer Theorem for Massive
Theories}, arXiv:1012.1454

\bibitem {B-S}Buchholz and S.J Summers, \textit{Scattering in Relativistic
Quantum Field Theory: Fundamental Concepts and Tools}, arXiv:math-ph/0405058

\bibitem {Buch}D. Buchholz, Commun. math. Phys. \textbf{36}, (1974) 243

\bibitem {B-M}J. Bros and J. Mund, \textit{Braid group statistics implies
scattering in three-dimensional local quantum physics}, arXiv:1112.5785

\bibitem {Wald}R. M. Wald, \textit{The Formulation of Quantum Field Theory in
Curved Spacetime}, arXiv:0907.0416

\bibitem {causal}B. Schroer, \textit{Causality and dispersion relations and
the role of the S-matrix in the ongoing research}, Foundations of Physics
\textbf{42}, (2012) 1481, arXiv:1107.1374

\bibitem {Bu-Yn}D. Buchholz and J. Yngvason, \textit{There are no causality
problems for Fermi's two atom system}, Physical Review Letters \textbf{73},
(1994) 613-616

\bibitem {PRL}G. C. Hegerfeldt, Causality problems in Fermi's two atom system,
Phys. Rev. Lett. \textbf{72}, (1994) 596-599

\bibitem {Duff}M. J. Duff, \textit{String and M-theory: answering the
critics}, arXiv:1112.0788

\bibitem {Smolin}L. Smolin, \textit{A perspective on the landscape problem}, arXiv:1202.3373

\bibitem {ABL}D. Buchholz, C. D'Antoni and R. Longo, \textit{Nuclear maps and
modular structures. I. General properties}, J. Funct. Anal. \textbf{88} (1990) 233

\bibitem {Quote}http://www.scribd.com/doc/13093631/Einstein-in-His-Own-Words

\bibitem {Mal}J. Maldacena, \textit{The Illusion of Gravity}, Scientific
American, (November 2005) page 56-63

\bibitem {Fub}P. Di Vecchia, \textit{The birth of string theory}, Lect. Notes
Phys.\textbf{737}, (2008) 59-118, arXiv 0704.0101

\bibitem {Jones}Vaughan F. R. Jones and V. S. Sunder, \textit{Introduction to
subfactors}, Cambridge University Press (1997) Volume 234

\bibitem {DR}S. Doplicher and J. E. Roberts, \textit{Why there is a field
algebra with a compact gauge group describing the superselection structure in
particle physics}, Commun. Math. Phys. \textbf{131}, (1990) 51-107

\bibitem {blog}http://golem.ph.utexas.edu/\symbol{126}distler/blog/archives/000987.html

\bibitem {Reh}K-H. Rehren, \textit{Algebraic Holography}, Annales Henri
Poincare\textbf{1}, (2000) 607,\ arXiv:hep-th/9905179

\bibitem {H-S}R. Haag and B. Schroer, \textit{Postulates of Quantum Field
Theory}, J. Mat. Phys. \textbf{3}, (1962) 248

\bibitem {H-Sw}R. Haag and J. A. Swieca, \textit{When does a quantum field
theory describe particles?}, Math. Phys. \textbf{1}, (1965) 308

\bibitem {Teg}M. Tegmark, \textit{Shut up and calculate}, arXiv:0709.4024, and
prior similar bizarre contributions

\bibitem {Smol}M. Arnsdorf and L. Smolin, \textit{The Maldacena Conjecture and
Rehren Duality}, arXiv:hep-th/0106073

\bibitem {Kay}B. S. Kay and L. Ortiz, \textit{Brick Walls and AdS/CFT}, arXiv:1111.6429

\bibitem {'t Hooft}G. 't Hooft, \textit{On the quantum structure of a black
hole}, Nucl. Phys. B \textbf{256}, (1985) 727

\bibitem {Swieca}B. Schroer, \textit{particle physics in the 60s and 70s and
the legacy of contributions \ by J. A. Swieca}, Eur.Phys.J.H \textbf{35},
(2010) 53, arXiv:0712.0371

\bibitem {L-S}J. H. Lowenstein and B. Schroer, Phys. Rev. \textbf{D7}, (1975) 1929

\bibitem {Lisa}L. Randall and R. Sundrum, Phys. Rev. Lett. \textbf{83}, (1999) 3370
\end{thebibliography}
\end{document}